\documentclass[a4paper,11pt]{article}
\pdfoutput=1
\usepackage{jheppub}
\usepackage[T1]{fontenc}
\usepackage{amsfonts}
\usepackage{amsmath}
\usepackage{amssymb}
\usepackage{multirow}
\usepackage{graphicx}
\usepackage{mathtools}
\usepackage{comment}
\usepackage{subcaption}

\newcommand{\M}{{\mathcal M}}
\renewcommand{\O}{{\cal O}}

\def\be{\begin{equation}}
\def\ee{\end{equation}}
\def\bea{\begin{eqnarray}}
\def\eea{\end{eqnarray}}

\newcommand{\tr}{{\rm {tr}}}

\newcommand{\Tr}{{\rm {Tr}}}

\numberwithin{equation}{section}

\begin{document}

\title{Bootstrapping supersymmetric (matrix) quantum mechanics}

\author[a, b]{Samuel Laliberte,}
\emailAdd{samuel.laliberte@oist.jp}
\author[b, c]{Brian McPeak}
\emailAdd{bmmcpeak@syr.edu}
\affiliation[a]{Okinawa Institute of Science and Technology Graduate University, 1919-1, Tancha, Onna, Kunigami District, Okinawa 904-0495, Japan.}
\affiliation[b]{McGill University, Montr\'{e}al, QC, H3A 2T8, Canada}
\affiliation[c]{Syracuse University, Syracuse, NY, 13244, USA}

\date{\today}

\abstract{We apply the quantum-mechanics bootstrap to supersymmetric quantum mechanics (SUSY QM) and to its matrix relative, the Marinari–Parisi model, which is conjectured to describe the worldvolume of unstable $D0$ branes. Using positivity of moment matrices together with Heisenberg, gauge, and (zero-temperature) thermal constraints, we obtain rigorous bounds on ground-state data. In the cases where SUSY is spontaneously broken, we find bounds that apply to the lowest-energy normalizable eigenstate.

For $N = 1$ SUSY QM with a cubic superpotential, we obtain tight bounds that agree well with available approximation methods. At weak coupling they match well with the semiclassical instanton contribution to SUSY-breaking ground-state energy, while at strong coupling they exhibit the expected scaling and match well with Hamiltonian truncation.

For the SUSY matrix QM, we construct a $44 \times 44$ bootstrap matrix and obtain bounds at large $N$. At strong coupling, we obtain the expected $E \sim \kappa g^{2/3}$ scaling of $E$ with $g$ and extract a lower bound on the coefficient $\kappa > .196$. At small coupling, the theory has a critical point $g_c$ where the two wells merge into one. We find a spurious kink at $g = \sqrt{2} g_c$. We attribute this to truncation error and solver limitations, and discuss possible improvements.

}
\maketitle

\section{Introduction}

Matrix models occupy a special place in the space of quantum mechanical models. They are flexible enough to describe a wide array of physical systems, across high energy theory and condensed matter physics. And yet they are often structured enough to yield to exact or approximate solutions even at strong coupling. Their noncommutative structure, in particular, makes them a rich playground for quantum gravity, and a significant amount of insight into 2d quantum gravity, string theory, and M-theory has been gained by studying them. An early study into matrix quantum mechanics \cite{Brezin:1977sv} was inspired by trying to find solvable systems in the planar (large $N$) limit. The fact that matrix models at large $N$ can describe random 2d surfaces, anticipated by 't Hooft \cite{tHooft:1973alw} implies a deep connection with string theory, and the topic experienced a renaissance in the late 80s and early 90s -- see \cite{Klebanov:1991qa, Ginsparg:1993is, Polchinski:1994mb} for early reviews or \cite{Anninos:2020ccj} for a more recent one. These efforts led to exact solutions for strings in two or fewer target space dimensions and provided an example of what a non-perturbative definition of string theory might look like. 

A particularly important class of 1d matrix models are those describing the dynamics of D0 branes, whose infinite-$N$ limit is conjectured to describe $M$-theory \cite{Banks:1996vh, Susskind:1997cw}. This model, now called the BFSS model, was derived \cite{deWit:1988wri} as a regularized description of a single supersymmetric membrane, and discussed even earlier as a reduction of 10d $N = 1$ SYM \cite{Baake:1984ie, Flume:1984mn, Claudson:1984th}. Recently this model, and some of its generalizations \cite{Ishibashi:1996xs, Berenstein:2002jq} have enjoyed renewed attention \cite{Herderschee:2023pza, Herderschee:2023bnc, Komatsu:2024vnb, Komatsu:2024bop, Komatsu:2024ydh, Lin:2023owt, Lin:2024vvg}, including using the bootstrap methods that will be the subject of this paper.

Supersymmetry is a key aspect of the BFSS model. It constrains its Hamiltonian to a very simple set of three terms with one coupling, and provides constraints that are essential to extracting bootstrap bounds. However the large number of matrices (9 bosonic plus 16 fermionic) poses a significant practical challenge, and developing efficient methods is an active topic. In this paper, we will be interested in a much simpler class of supersymmetric matrix models with a single bosonic matrix whose Hamiltonian is defined with an arbitrary superpotential $W$. These theories were first considered when Marinari and Parisi \cite{Marinari:1990jc} showed that the case of a cubic superpotential arises as the continuum limit as a sum over triangulations of a string in 2d superspace.  
\cite{Dabholkar:1991te} showed that the instanton effects responsible for SUSY breaking show the $e^{-C/g_s}$ scaling discovered for closed strings by Shenker \cite{Shenker:1990uf,Ferretti:1993fu}. The collective field theory approach was developed in \cite{Cohn:1992zj, Rodrigues:1992by, Brustein:1993nc}. 
The nature of the continuum limit was discussed in \cite{Chaudhuri:1994yk} and the ungauged model was discussed in \cite{Kim:2005rp}. For a while the string theory describing the model was unknown: in \cite{McGreevy:2003dn} it was argued that the model describes unstable D0 branes in type IIB string theory. The model with logarithmic superpotential to strings in AdS${}_2$ \cite{Verlinde:2004gt}.

The present paper initiates a numerical study of this model. After briefly reviewing the bootstrap and supersymmetric quantum mechanics, we first derive constraints for pure (non-matrix) quantum mechanics. We find fairly tight bounds and are able to uncover a number of features of the theory, such as the degeneracy of the spectrum and the appropriate behaviors at large and small coupling. We are able to check our results against perturbation theory, the WKB approximation, and Hamiltonian truncation in various limits, and find good agreement. 

We then turn to the matrix theory, where there are fewer known results. We find that tight bounds are harder to obtain -- this is essentially due to the existence of double traces, which both make the computations less tractable, and introduce more variables into the optimization problems, leading to weaker bounds. We find the expected scaling at large $N$ and large coupling, though the lower bound on the ground state energy is somewhat lower than what is suggested by the WKB approximation, raising the question of how close our answer is to the true ground state. We also examine the system at weak coupling, where it is expected to have a critical point in a particular double-scaled limit. Interestingly, we find that the lower-bound zero until $g = \sqrt{2} g_c$, at which point it increases quadratically\footnote{It is tempting to think that we have simply normalized the Hamiltonian in such a way that the critical point is larger by a factor of $\sqrt 2$. This can't be the case due to the scaling at the critical point -- our bounds point to quadratic scaling at the kink while the expected behavior is $E \sim (g - g_c)^{5/2}$. If our kink we at the critical point, our lower bound would be above the true value.}. We believe that this is a truncation artifact and hope that future algorithmic improvements will allow larger bootstrap matrices and stronger bounds to resolve the issue.

Along the way, we implemented a non-linear solver using the method discussed in \cite{Han:2020bkb}, in addition to using Mathematica's SDP solver and SDPB \cite{Simmons-Duffin:2015qma}. We found that for practical purposes, it was always preferable to use the latter two methods, which are only suited to semi-definite problems -- problems where the constraints can be expressed as positive-semidefiniteness of matrices that are linear in the constraints. The reason is that while the semi-definite solvers do not allow for the addition of non-linear constraints (such as those that arise due to large $N$ factorization), but they are so much faster that larger bootstrap matrices can be used, allowing stronger bounds for a given amount of computing time. However, there are some applications where non-linear constraints are essential. For instance we show plots in the next section with non-convex regions -- these can only be made using non-convex constraints.

\paragraph{Quantum mechanics bootstrap}

Bootstrap, describing the array of techniques for constraining or solving physical systems using only their symmetries and positivity properties, has recently found a number of applications in quantum mechanics. These methods were inspired by well-developed bootstrap techniques for studying conformal field theories~\cite{Belavin:1984vu, Rattazzi:2008pe,Kos:2016ysd,Simmons-Duffin:2015qma, Poland:2018epd} which have also been applied to study S-matrices~\cite{Paulos:2016but, Paulos:2016fap, Paulos:2017fhb} and EFTs~\cite{Caron-Huot:2020cmc,Tolley:2020gtv, Sinha:2020win, Arkani-Hamed:2020blm}. One of its most appealing properties is the ability to obtain \textit{rigorous bounds} on various physical data (correlation functions, couplings, expectation values, etc.) which in some cases appear to converge to the true value as the computational power is increased. A similar set of techniques was uncovered for putting rigorous bounds on expectation values of operators in quantum systems. These were first explored in \cite{Han:2020bkb}, inspired by earlier work on matrix integrals~\cite{Lin:2020mme} (see also \cite{Koch:2021yeb,Kazakov:2021lel}). Already in~\cite{Han:2020bkb} it was appreciated that the methods are quite suitable for models of matrix quantum mechanics which are not exactly solvable. Even more remarkably, the method's complexity is independent of the matrix size $N$, which enters the algorithm as an easily tunable parameter.

Since its discovery, quantum mechanical bootstrap and related methods have been used for purposes including constraining ground-state physics and higher spectra in quantum mechanics~\cite{Berenstein:2021dyf,Berenstein:2021loy,Nancarrow:2022wdr,Aikawa:2021eai,Tchoumakov:2021mnh,Aikawa:2021qbl,Du:2021hfw,Bai:2022yfv,Hu:2022keu,Berenstein:2022ygg,Guo:2023gfi,Blacker:2022szo,Sword:2024gvv,Huang:2025sua, Lin:2025srf}, integrable systems \cite{Aikawa:2025dvt}, other theories of matrix quantum mechanics~\cite{Lin:2023owt,Lin:2024vvg}, scattering \cite{Berenstein:2023ppj}, open systems \cite{Robichon:2024apx}, lattice systems~\cite{PhysRevLett.108.200404,Anderson:2016rcw,Lawrence:2021msm,Lawrence:2022vsb,Kazakov:2022xuh,Cho:2022lcj, Cho:2023ulr, Kazakov:2024ool, Li:2024wrd,Guo:2025fii}, systems at finite temperature~\cite{Fawzi:2023fpg,Cho:2024kxn} and with many degrees of freedom \cite{Kull:2022wof,Wang:2023hss, Cho:2024owx, Scheer:2024eyu,Berenstein:2024ebf,Jansen:2025nlc, Gao2025}. Constraints on real time evolution have also been considered in \cite{Lawrence:2024hjm, Lawrence:2024mnj} as has out-of-equilibrium physics in \cite{Cho:2025dgc, Cho:2025nlv}. Issues related to the algorithms / methods have been considered in \cite{Berenstein:2022unr,Berenstein:2025itw}. A pedagogical introduction to the methods was recently given in \cite{Lin:2025iir}.

The version of the method we will employ works as follows. We consider a quantum system with a Hamiltonian $H$ and a Hilbert space $\mathcal H$. Positivity of norms on the Hilbert space implies that 
\begin{align}
    \langle \Psi | \Psi \rangle > 0 \, .
\end{align}
This statement can be repackaged as a statement about positive matrices as follows: first consider any eigenstate of $H$, $| E \rangle$, and consider a finite set of operators $\mathcal{O}_1$, $\mathcal{O}_2$, $\mathcal{O}_3, ... $, $\mathcal{O}_n$. We can form different quantum states by acting on $|E \rangle$ with these operators,
\begin{align}
    | \Psi (c_1, c_2, ..., c_n) \rangle \ = \ \sum c_i \mathcal{O}_i |E \rangle \, .
\end{align}
Positivity of 
\begin{align}
    \langle \Psi (c_1, c_2, ..., c_n)  | \Psi (c_1, c_2, ..., c_n) \rangle
\end{align}
for any values of $c_1$, $c_2, ...$ is equivalent to the statement that the matrix
\begin{align}
    \M = \begin{pmatrix}
        \langle \mathcal{O}_1 \mathcal{O}_1 \rangle && \langle \mathcal{O}_1 \mathcal{O}_2 \rangle && \langle \mathcal{O}_1 \mathcal{O}_3 \rangle && ... \\
        \langle \mathcal{O}_2 \mathcal{O}_1 \rangle && \langle \mathcal{O}_2 \mathcal{O}_2 \rangle && \langle \mathcal{O}_2 \mathcal{O}_3 \rangle && ... \\
        \langle \mathcal{O}_3 \mathcal{O}_1 \rangle && \langle \mathcal{O}_3 \mathcal{O}_2 \rangle && \langle \mathcal{O}_3 \mathcal{O}_3 \rangle && ... \\
        ... & ... & ... & ...
    \end{pmatrix}
\end{align}
is positive definite, $\M \succ 0$. Here and for the rest of the paper, we take $\langle \mathcal{O} \rangle$ to be the expectation value of $\mathcal{O}$ \textit{in an eigenstate of the Hamiltonian}. See \cite{Lawrence:2024mnj} for what happens when one is not in an eigenstate -- naturally, the state will evolve, and a rigorous method exists for finding bounds in that case.

Now, being in an eigenstate of $H$ has another nice effect -- it implies the following equation:
\begin{align}
    \langle [ H , \mathcal{O}] \rangle = 0 \, .
    \label{eq:H_constraint}
\end{align}
This is what we will refer to as the ``Heisenberg equation'' or simply as the equation of motion. This equation is crucial: \textit{a priori} the positivity of $\M$ has very limited implications for the expectation values of the operators because the matrix $\M$ has too many independent entries. However, the Heisenberg equation serves to relate many of the elements of the matrix $\M$, allowing tight bounds to be found.

If we want $\M$ to be a finite matrix, we need to choose a finite set of operators -- we use $m$ to be the number of operators. One important point is that if one adds an operator, forming a new $(m+1)\times(m+1)$ matrix, then the bounds obtained should be strictly equal to or stronger than the previous bounds. This follows directly from the fact that a positive definite matrix has all positive definite principal minors, so the size-$m$ positivity constraints are implied by the $m+1$ constraints. 

By imposing positivity plus the Heisenberg equation, it is possible to get rigorous bounds on any desired expectation values. Note that these constraints are convex (a sum of solutions is a solution) so the bounds will form convex regions that are essentially the convex hull of the allowed values. In practice one is often interested in the ground state properties like the energy. There is good evidence in different settings that as $m$ is increased, the lower bound on the energy converges to the true ground state energy\footnote{It is possible, by using a non-convex constraint $\langle H \mathcal{O} \rangle = \langle H \rangle \langle \mathcal{O} \rangle$, to find bounds which converge to the true values in individual eigenstates -- this essentially gives a series of small islands around the allowed values. However efficient implementation, especially important when there are many relevant operators, relies on algorithms specific to convex problems, so this non-convex constraint is sometimes ignored.}.

\section{Supersymmetric quantum mechanics}

Supersymmetric quantum mechanics is a rich topic with deep connections to mathematics. A nice introduction to the subject is given in \cite{Tong}. We will start with a brief review of the aspects needed for our bootstrap analysis.

The supersymmetry algebra for quantum mechanical systems includes a supercharge $Q$ and its conjugate $Q^\dagger$, satisfying the relations
\begin{align}
    \label{eq:alg}
    \left\{ Q, Q^\dagger \right\} = 2 H\, , \qquad Q^2 = 0
\end{align}
An operator $O^\dagger O$ will have non-negative eigenvalues. $H$ is a sum of such operators so we have $E \geq 0$. The Hilbert space of a supersymmetric theory can be decomposed into a bosonic and a fermionic sector, $\mathcal{H} = \mathcal{H}_B \oplus \mathcal{H}_F$. A state $|0 \rangle \in \mathcal{H}_B$ has the properties that 
\begin{align}
    Q | 0 \rangle = 0 \, , \qquad Q^\dagger|0\rangle = |1\rangle \in \mathcal H_F \, .
\end{align}
So $Q^\dagger$ maps $\mathcal H_B$ to $\mathcal H_F$, and vice versa for $Q$. One can show that $Q$ and $Q^\dagger$ commute with the Hamiltonian; as a result, $|0\rangle$ and $|1\rangle$ have the same energy, and we see that supersymmetric eigenstates come in pairs -- one bosonic and one fermionic.

An exception to that statement is for states that are annihilated by both $Q$ and $Q^\dagger$. From the algebra we see that such states must have zero energy. So zero energy states are not necessarily paired like positive energy states. When 0 energy states do not exist, SUSY is said to be \textit{spontaneously broken} because the lowest energy state is not annihilated by the SUSY generators $Q$ and $Q^\dagger$.

\subsubsection*{Supersymmetric Hamiltonian from a superpotential}

The algebra in equation~\eqref{eq:alg} can be realized by a number of systems, including finite dimensional ones, but we shall be interested in the example of a supersymmetric particle on a line. In addition to the usual Hilbert space of functions on the line, we must add a finite (two-dimensional) degree of freedom to get a $\mathbb{Z}_2$ graded vector space. So the full Hilbert space is $L^2(\mathbb{R}) \otimes \mathbb{C}^2$. A convenient choice for the supercharges is 
\begin{align}
    Q = (p + i W'(x)) \psi \, ,\qquad Q^\dagger  = (p - i W'(x) ) \psi^\dagger
\end{align}
Here $W(x)$ is an arbitrary function called the superpotential and $\psi$ and $\psi^\dagger$ are fermionic raising and lowering operators acting on the $\mathbb{C}^2$ part of the Hilbert space. They obey the anti-commutation relations 
\begin{align}
	\{\psi^\dagger , \psi \} & = 1 \, ,\qquad 
	\{\psi^\dagger , \psi^\dagger \}  = 0 \, , \qquad 
	\{\psi , \psi \}  = 0 \, .
\end{align}
The algebra above has a two-dimensional irreducible representation spanned by the states $|0 \rangle$ and $| 1 \rangle$ with the properties that
\be
\psi | 0 \rangle = 0 \quad \text{and} \quad |1 \rangle = \psi^\dagger |0 \rangle \, .
\ee
In this basis $\psi$ and $\psi^\dagger$ can be represented as
\begin{align}
    \psi = \begin{pmatrix}
        0 & 0 \\ 1 & 0
    \end{pmatrix} \, , \qquad \psi^\dagger = \begin{pmatrix}
        0 & 1 \\ 0 & 0
    \end{pmatrix} 
\end{align}
The Hamiltonian can be computed from the algebra and is 
\begin{align}
    H = \frac{1}{2} p^2 + \frac{1}{2} W'(x)^2 + \frac{1}{2} [\psi^\dagger, \psi] \,  W''(x) \, .
\end{align}
We see that there is a family of supersymmetric theories parametrized by the choice of superpotential $W$.

These theories have another symmetry related to the conserved fermion number
\begin{align}
    F = \psi^\dagger \psi \, .
\end{align}
which satisfies $F | 0 \rangle = 0$ and $F |1 \rangle = |1\rangle$. So $F$ has eigenvalue $0$ on the bosonic part of the Hilbert space and eigenvalue 1 on the fermionic part. It is useful to note that one can study a particular sector of the theory by projecting onto $\mathcal H_B$ or $\mathcal{H}_F$. For the former, $[\psi^\dagger, \psi] = -1$ and for the latter it is $1$. In each sector the Hamiltonian is one-dimensional and can be written as 
\begin{align}
    H = \frac{1}{2} p^2 + \frac{1}{2} W'(x)^2 + \frac{1}{2} \epsilon \,  W''(x)
    \label{eq:H_eff}
\end{align}
where $\epsilon = 2F - 1 = \pm 1$ for fermionic / bosonic states, respectively.

\subsubsection*{Ground states and SUSY breaking}

Recall the $E = 0$ states should be annihilated by both $Q$ and $Q^\dagger$. For the particle in $\mathbb R$, a general wavefunction
\begin{align}
    \Psi(x) = \begin{pmatrix}
        \psi_B \\ \psi_F
    \end{pmatrix}
\end{align}
satisfies the equations
\begin{align}
    \psi_B ' + W' \psi_B = 0 \, ,\qquad \psi_F ' - W' \psi_F = 0
\end{align}
The solutions are $\psi_B = \exp(-W)$ and $\psi_F = \exp(W)$. 

One immediate result of this is that there is no normalizable ground state for polynomial superpotentials where the leading power is odd. For even superpotentials, we do have normalizable ground states, though not for $\psi_B$ and $\psi_F$ at the same time -- one of them has to be zero. We will see below that the existence of normalizable ground states can be studied using the bootstrap. Imposing normalizability -- that $\langle 1 \rangle = 1$ in an eigenstate -- will single out the normalizable states. Lower bounds on the energy, for instance, would become lower bounds on the energy in normalizable states.

\subsection{Bootstrap constraints from SUSY}
\label{sec:qm_bootstrap}

To obtain bounds, we consider a matrix $\mathcal{M}$ of the form
\be
\mathcal{M}_{ij} = \langle x^{i+j} \rangle \, .
\ee
and find a recursion relation relating expectation values of the form $\langle x^{p} \rangle$ to each other -- the result is that $\mathcal M$, no matter how large, will depend on a finite set of $\langle x^{p} \rangle$. We then impose that $\mathcal{M}$ is positive semidefinite to obtain bounds on the energy and the allowed values for the $\langle x^{p} \rangle$ expectation values.

Let us first see how this can be done for an arbitrary superpotential. We shall need to use $\langle [ H , \mathcal{O}] \rangle = 0$ for the cases $\mathcal{O} = x^{t+1}$ and $\mathcal{O} = x^t p$. Making use of these two cases, along with the identity $[x^t , p] = i t x^{t-1}$, we obtain
\be
4 t \langle x^{t-1} p^2 \rangle = - t(t-1)(t-2) \langle x^{t-3} \rangle + 4 \langle x^t W'' W' \rangle + 2 \langle \epsilon x^t W''' \rangle \, .
\label{eq:rec1}
\ee
We then relate the quantity above to the energy of the system $E$ by making use of the constraint $\langle \mathcal{O} H \rangle = E \langle \mathcal{O} \rangle$, which is a stronger version of the Heisenberg equation. Here, making use of $\mathcal{O} = x^{t-1}$, we obtain
\be
\langle x^{t-1} p^2 \rangle = 2 E \langle x^{t-1} \rangle - \langle x^{t-1} W'^2 \rangle - \epsilon \langle x^{t-1} W'' \rangle \, .
\label{eq:rec2}
\ee
Finally, combining equation \ref{eq:rec1} and \ref{eq:rec2}, we obtain the recursion relation
\begin{align}
    8 t E \langle x^{t-1} \rangle + t(t-1)(t-2) \langle x^{t-3} \rangle - 4 t \langle x^{t-1} W'^2 \rangle - 4 t \epsilon \langle x^{t-1} W'' \rangle \nonumber \\
    \quad - 4 \langle x^t W'' W' \rangle - 2 \epsilon \langle x^t W''' \rangle = 0 ,
    \label{eq:reqanyW}
\end{align}
which relates the energy $E$ to correlators involving functions of $x$. 
This recursion relation is not always solvable, but for the cases of interest here -- quadratic and cubic superpotentials, it will give rise to useful constraints.

\subsubsection{SUSY harmonic oscillator}

The simplest case to consider is a system with a superpotential given by 
\be
W = \frac{1}{2} \omega x^2 \, ,
\ee
which gives rise to an Hamiltonian of the form
\be
H = \frac{1}{2} p^2 + \frac{1}{2} \omega^2 x^2  + \frac{1}{2} \epsilon \omega \, .
\ee
This system can be solved exactly using conventional quantum mechanics methods. We obtain the energy levels
\be
E = |\omega| \left( n + \frac{1}{2} \right) + \frac{1}{2} \epsilon \omega
\ee
where $n \in \mathbf{N}$. The spectrum above has two interesting regimes : $\omega > 0$ and $\omega < 0$. When $\omega > 0$, we obtain the spectrum
\be
E_{\epsilon = - 1} =  |\omega| n \quad , \quad E_{\epsilon = 1} =  |\omega| (n + 1)
\ee
When $\omega < 0$, the $E_{\epsilon = - 1}$ and $E_{\epsilon = 1}$ spectra are interchanged and we obtain
\be
E_{\epsilon = - 1} =  |\omega| (n + 1) \quad , \quad E_{\epsilon = 1} =  |\omega| n
\ee
In both cases, the system has a unique $E = 0$ ground state, and two degenerate states for each energy level when $E > 0$. When $\omega > 0$, this unique ground state is found when $n = 0$ and $\epsilon = - 1$. When $\omega < 0$, this ground state is found when $n = 0$ and $\epsilon = 1$.

\begin{figure}[h]
	\begin{subfigure}{.5\textwidth}
		\centering
		\includegraphics[width = 7.0cm]{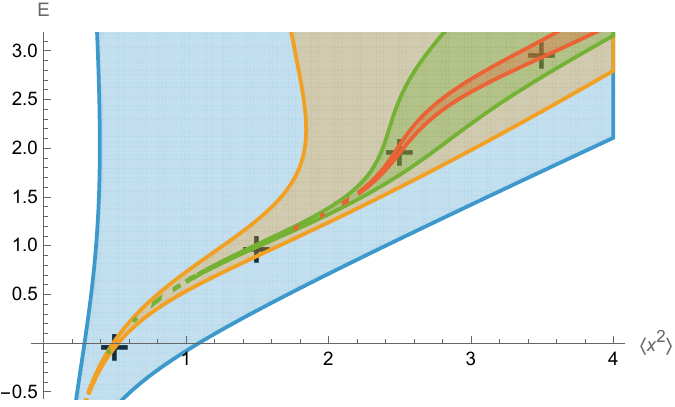}
	\end{subfigure}
	\begin{subfigure}{.5\textwidth}
		\centering
		\includegraphics[width = 7.0cm]{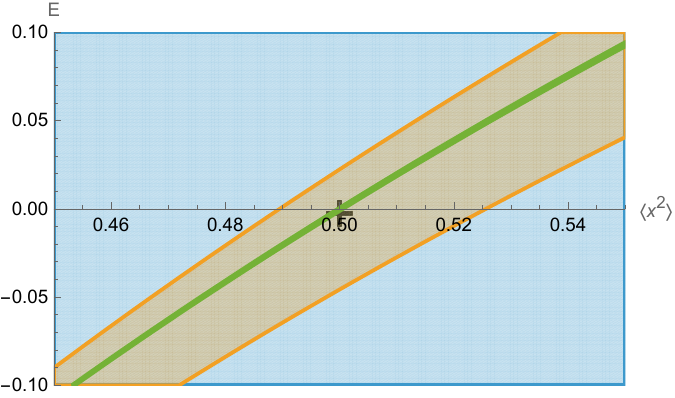}
	\end{subfigure}
    \caption{Quadratic superpotential with $\epsilon = -1$, $\omega = 1$. The matrix sizes are $K \times K$ with $K = 4$ (blue), $K = 8$ (orange), $K = 12$ (green) and $K = 16$ (red). The exact results are represented by black crosses. The results for $\epsilon = 1$ are the same as all of the energies shifted up by 1. Thus all states except the ground state are paired. These constraints use $\langle H \mathcal O \rangle = E \langle \mathcal O \rangle$}, hence the non-convex regions.
    \label{fig:quadratic_constraints}
\end{figure}

These energy levels can be found from the bootstrap approach developed in the last section. For the present case, the recursion relation reads
\be
(t-3)(t-2)(t-1) \langle x^{t-4} \rangle + 4 (t - 1) ( 2 E - \epsilon \omega) \langle x^{t-2} \rangle - 4 t \omega^2  \langle x^{t} \rangle = 0 \, .
\ee
The constraints found by imposing positivity and the recursion relation above can be found in figure \ref{fig:quadratic_constraints}. As we can see, for $\omega = 1$, the constraints are consistent with the exact spectrum.

\subsubsection{Quartic correction to the SUSY harmonic oscillator}
\label{sec:quartic_corrections}

\begin{figure}[h]
	\begin{subfigure}[t]{.5\textwidth}
		\centering
		\includegraphics[width = 7.5cm]{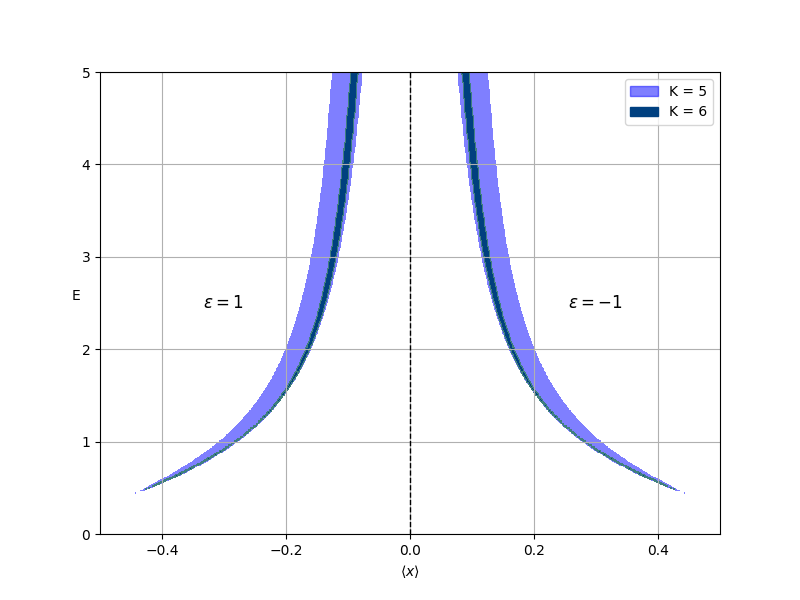}
	\end{subfigure}
	\begin{subfigure}[t]{.5\textwidth}
		\centering
		\includegraphics[width = 7.5cm]{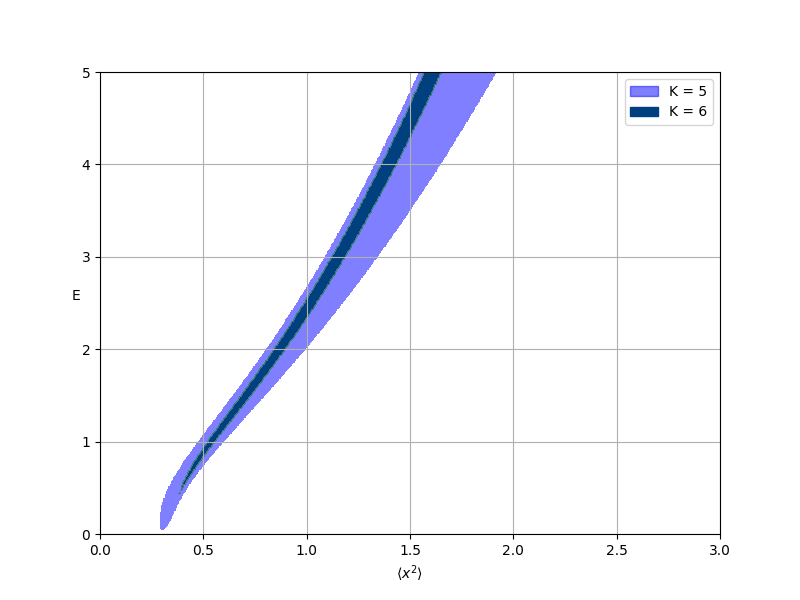}
	\end{subfigure}
    \caption{(Left) Constraint region for E vs $\langle x \rangle$ for a bootstrap matrix of size K x K where K is 5 (blue) and 6 (dark blue). The $\langle x \rangle < 0$ portion of the figure shows the results when $\epsilon = 1$ and the $\langle x \rangle > 0$ portion of the figure shows the results when $\epsilon = -1$. (Right) Constraint region for E vs $\langle x^2 \rangle$ for a bootstrap matrix of size K x K where K is 4 (blue) and 5 (dark blue).}
    \label{fig:cubic_constraints}
\end{figure}

We now add a correction to the harmonic oscillator in the form of a quartic term, and study the impact on the bounds found in the previous section. For the present supersymmetric system, one can achieve this task by considering a superpotential of the form
\be
W = \frac{1}{\sqrt{2} g} x + \frac{g}{3\sqrt{2}} x^3 \, .
\ee
In this case, the Hamiltonian can be written in the form
\be
H = \frac{1}{2} p^2 + \frac{1}{4g^2} - \frac{g^2}{4} + \frac{1}{2} \left( x + \frac{1}{\sqrt{2}} \epsilon g \right)^2 + \frac{1}{4} g^2 x^4 \, .
\ee
which makes explicit the fact that this system describes a harmonic oscillator with a minimum located at $x_{min} = - \frac{1}{\sqrt{2}} \epsilon g$, and an off-center quartic correction. To constrain this system, we can use the recursion relations 
\begin{align}
	(t-3)(t-4)(t-5) \langle x^{t-6} \rangle + 2 \left( 4 E - \frac{1}{g^2} \right) (t-3) \langle x^{t-4} \rangle \nonumber \\
	\quad - 2 \sqrt{2} \epsilon g (2t-5) \langle x^{t-3} \rangle - 4 (t-2) \langle x^{t-2}\rangle - 2 g^2 (t-1) \langle x^{t} \rangle = 0 
    \label{eq:cubic_qm_constraints} \, 
\end{align} 
found from substituting the present superpotential in Equation \ref{eq:reqanyW}. In this case, imposing that $\mathcal{M}_{ij} = \langle x^{i+j} \rangle$ is positive semi-definite given the constraints of equation \ref{eq:cubic_qm_constraints} leads to the constraint region shown in Figure \ref{fig:cubic_constraints} when $g = 1$. 
As we can see, the constraint region has a $\textbf{Z}_2$ symmetry around the $x = 0$ axis. This symmetry reflects the degeneracy of the excited states in the system due to supersymmetry. Moreover, the bounds hint at the existence of a degenerate minimum of the energy greater than zero. This suggests that supersymmetry may be broken by the quartic correction, which is expected to happen for every quantum mechanical system with a superpotential of odd power in $x$ such as the present one. 

\begin{figure}[h]
    \centering
    \includegraphics[width = 12.0cm]{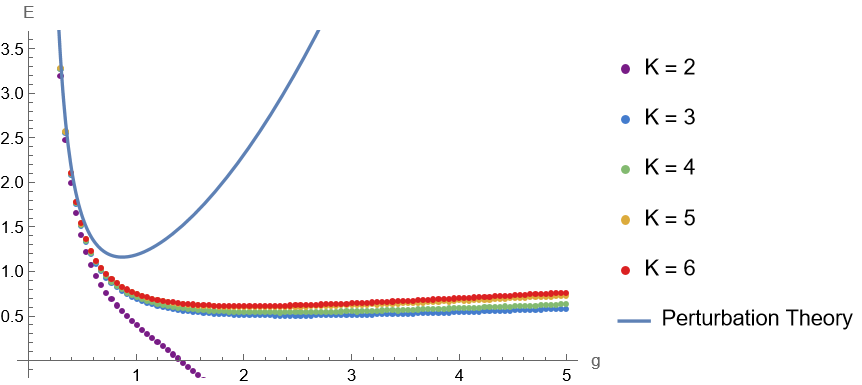}
    \caption{Bounds on the minimum of the energy vs g (colored dots) compared to expectation from perturbation theory (solid blue line). The bounds were found using bootstrap matrices of the form shown in Equation \ref{eq:bloc_M}, where let the size $K$ of the $K$ x $K$ blocks $A$, $B$, $C$, and $D$ take the values 2 (purple dots), 3 (blue dots), 4 (green dots), 5 (yellow dots), and 6 (red dots).}
    \label{fig:Emin_vs_g_cubic_qm}
\end{figure}

To probe the broken supersymmetry in greater detail, one can push the present analysis further and try to find the bottom of the constraint region described in Figure \ref{fig:cubic_constraints} for different values of the coupling g. This turns out to be quite difficult if we try to use the constraint equations \ref{eq:cubic_qm_constraints} because of terms of the form $E \langle O \rangle$, which makes the present semi-definite programming problem non-convex. However, it's possible to overcome this issue by modifying our approach slightly to only use constraints that can be found from $\langle [ H , O ] \rangle = 0$. In the present case, our strategy was to consider a bootstrap matrix with the block form
\begin{equation}
    \mathcal{M} = 
    \begin{pmatrix}
    A & B \\
    C & D
    \end{pmatrix} ,
    \label{eq:bloc_M}
\end{equation}
where the entries of each block are given by
\begin{align}
    A_{ij} & = \langle p^{i+j} \rangle & B_{ij} & = \langle p^i x^j \rangle \nonumber \\
    C_{ij} & = \langle x^{i} p^j \rangle & D_{ij} & = \langle x^{i+j} \rangle \, .
\end{align}
We then expressed $\langle H \rangle$ and $\mathcal{M}$ in terms of the family of correlators $\langle x^p \rangle$ such that $p \in \mathbf{N}$ using the Hamiltonian constraints $\langle [ H , O ] \rangle = 0$ for suitable values of $O$. A detailed explanation of how this can be achieved can be found in the Appendix \ref{ap:convex}. The minimum energy was then found by finding the parameters $\langle x^p \rangle$ such that $\langle H(\langle x^p \rangle)\rangle$ is minimized under the constraint that $\mathcal{M}(\langle x^p \rangle)$ must be positive semi-definite. This was done using readily available algorithms for solving convex semi-definite programming problems. In the present section, we show results found using Mathematica's built-in \texttt{SemidefiniteOptimization[]} function. This leads to the minimum of the energy found in Figure \ref{fig:Emin_vs_g_cubic_qm}.

As we can see in Figure \ref{fig:Emin_vs_g_cubic_qm}, a small ($K$ x $K$ blocks where $K$ = 2) bootstrap matrix allows for negative minimum energies. However, for $K \geq 3$, the bounds on the minimum start converging around a curve where $E_{min}$ is strictly positive. This matches our expectation that supersymmetry is spontaneously broken, and no zero-energy ground states are allowed for $g \geq 0$.

Quantitatively, we find that the minimum of the energy found from the bootstrap constraints matches expectations from perturbation theory in the $g \ll 1$ regime (see Appendix \ref{ap:pert_theory} for a detailed derivation of these results). As $g$ becomes greater than one, the bootstrap results start diverging from the results from perturbation theory. In this limit, the SemidefiniteOptimization[] function in Mathematica did not converge well. However, we observed that the bounds seem to grow in a way similar to the $N = 1$ Marinari-Parisi model in the large $g$ limit (see Figure \ref{fig:largegn1}). This is expected since for both models, the potential becomes dominated by a quartic term in the large $g$ limit. 

\subsubsection{$N = 1$ Marinari-Parisi model}

Finally let us comment on the model with
\begin{align}
    W = \frac{1}{2} x^2 + g \frac{1}{3} x^3 \, ,
\end{align}
which is the $N = 1$ limit of the original Marinari-Parisi model that will be studied in the next section. We bootstrapped this model using a set that includes all operators up to level 5, where $x$ has level 1 and $p$ has level 2 \cite{Lin:2024vvg}. The result is a $20 \times 20$ bootstrap matrix of all strings up to level 10. We also used the ``ground state constraints'' -- the methods of the thermal bootstrap \cite{Fawzi:2023fpg, Cho:2024owx} in the zero temperature limit, which requires that 
\begin{align}
    \langle \mathcal O_i^\dagger [H, \mathcal O_j] \rangle \succeq 0 \, .
\end{align}
The result, going again up to level 10, is an additional $11 \times 11$ matrix that must be positive in the ground state -- this will allow us to find upper bounds on the ground state energy and other expectation values. 

\begin{figure}[h]
	\begin{subfigure}{.5\textwidth}
		\centering
		\includegraphics[width = 7.0cm]{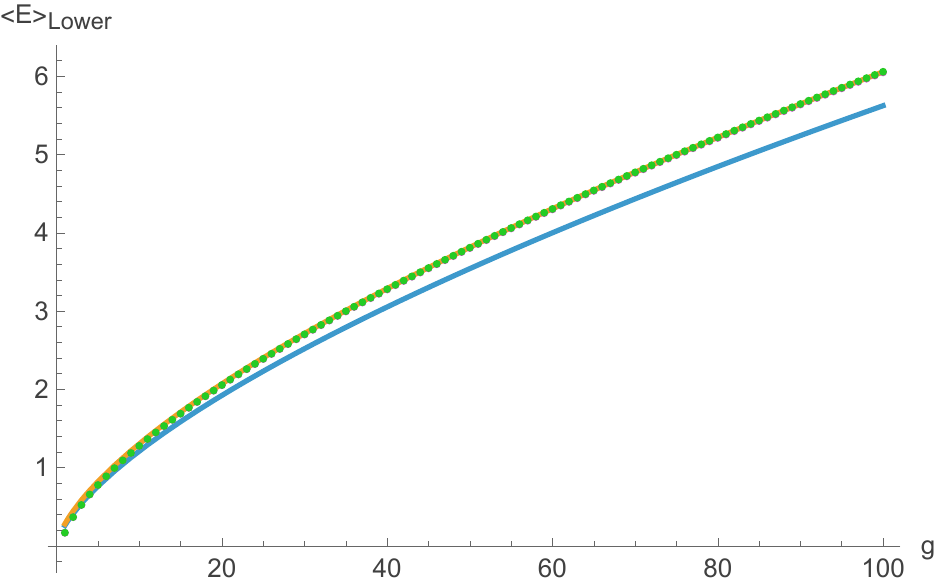}
	\end{subfigure}
	\begin{subfigure}{.5\textwidth}
		\centering
		\includegraphics[width = 7.0cm]{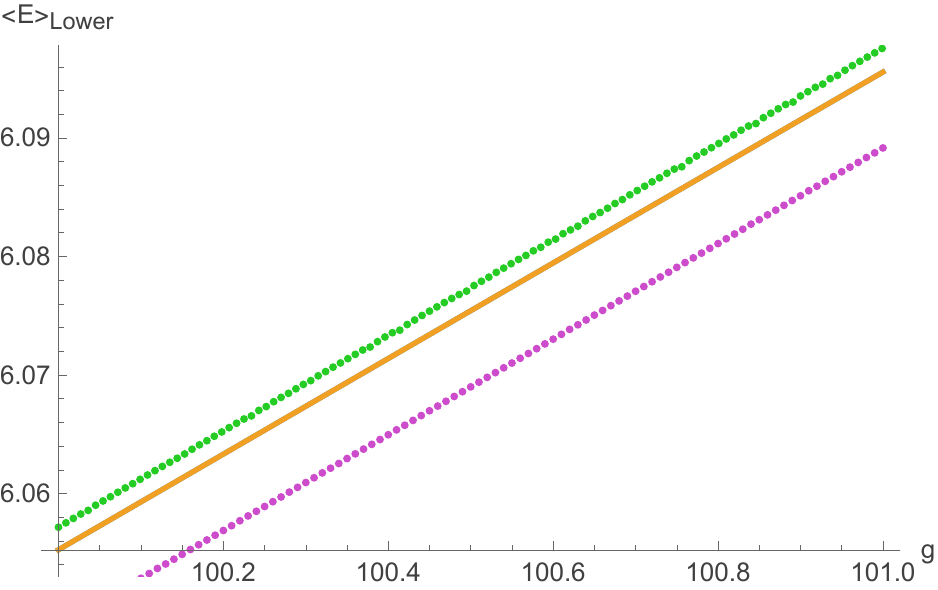}
	\end{subfigure}
    \caption{Ground state energy bounds for SUSY QM with $W = 1/2 x^2 + g/3 x^3$. The green dots are the upper bound and the purple are the lower bound. The blue line is the WKB approximation and the orange line is the approximation from the Rayleigh-Ritz method with a $60 \times 60$ matrix, using the leading $g^{2/3}$ part of the Hamiltonian~\eqref{eq:Hleading}.}
    \label{fig:largegn1}
\end{figure}

For large $g$, the results are given in figure~\ref{fig:largegn1}. The scaling at large $g$ can be demonstrated by a simple argument: rescaling $x \to g^{-1/3} x $ and $p \to g^{1/3} p$ leads to the Hamiltonian
\begin{align}
    \label{eq:Hleading}
    H = \frac{1}{2}g^{2/3} \left( p^2 + x^4 - 2x   + \mathcal O(g^{-2/3}) \right) \,.
\end{align}
At large $g$, the energy is proportional to $\kappa_0 \, g^{2/3}$, with some corrections at higher orders of $g^{-2/3}$. We have estimated $\kappa_0$ using the WKB approximation, which appears to be off by about $7 \%$, and the Rayleigh-Ritz (a.k.a. Hamiltonian truncation) method with a $60 \times 60$ matrix, which we find lies inside our allowed region.
\begin{figure}[h]
	\begin{subfigure}{.5\textwidth}
		\centering
		\includegraphics[width = 7.0cm]{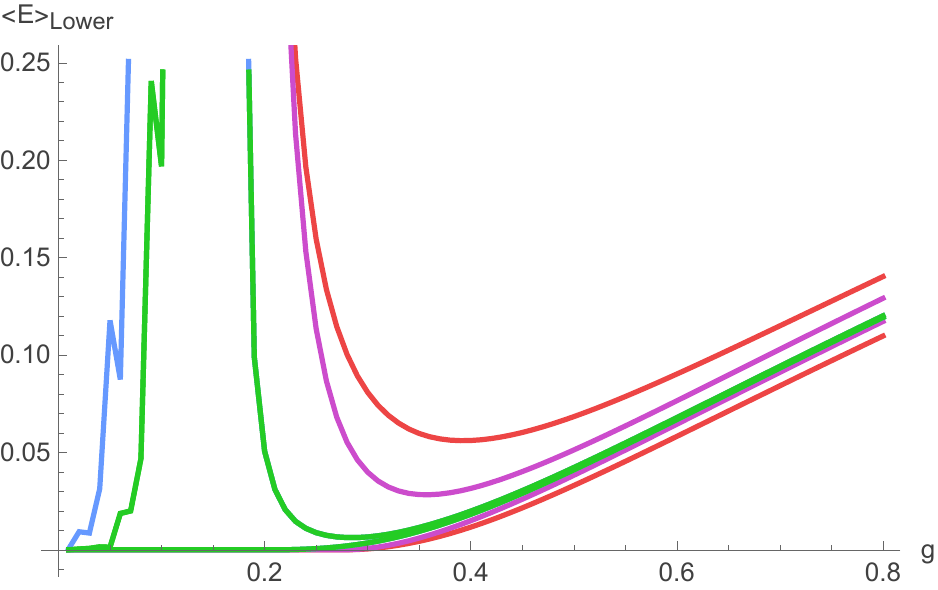}
	\end{subfigure}
	\begin{subfigure}{.5\textwidth}
		\centering
		\includegraphics[width = 7.0cm]{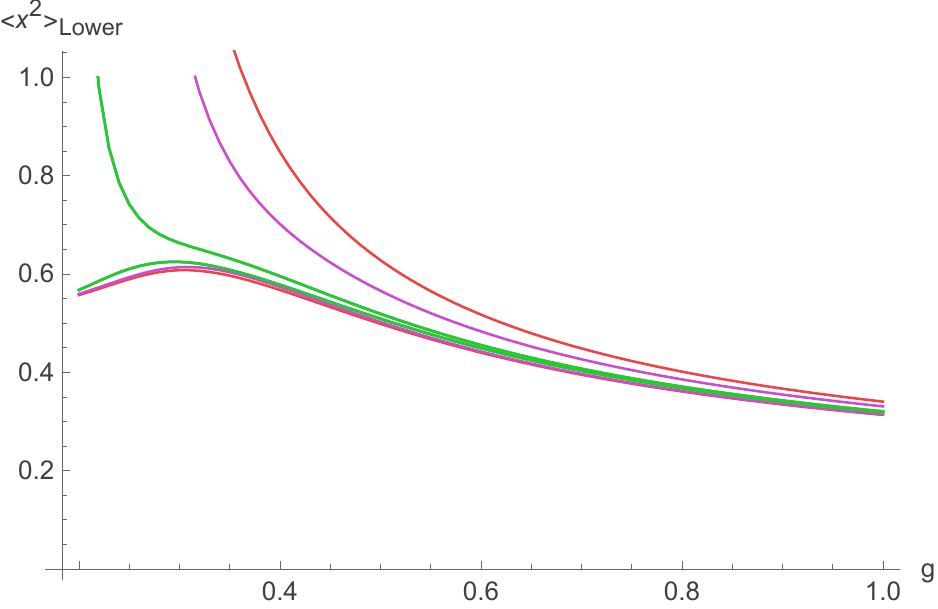}
	\end{subfigure}
    \caption{Ground state energy (left) and $x^2$ (right) lower bounds for SUSY QM with $W = 1/2 x^2 + g/3 x^3$ for level 7 (red), level 8 (purple), level 9 (blue) and level 10 (green). For smaller $g$ Mathematica's solver breaks down (see later figures).}
    \label{fig:smallgn1}
\end{figure}

\begin{figure}[h]
		\centering
		\includegraphics[width = 7.0cm]{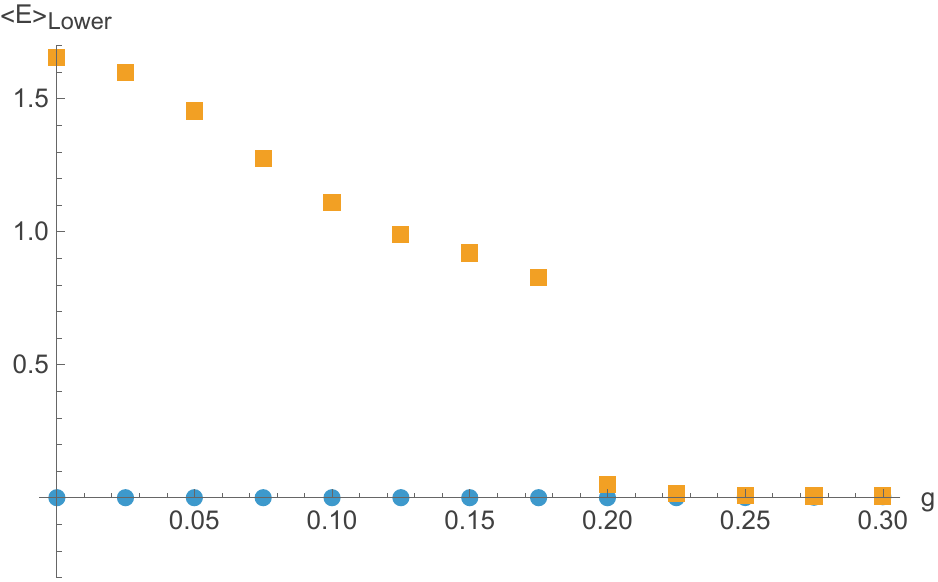}
    \caption{Ground state energy lower (blue) and upper (yellow) bounds using level 10 constraints. }
    \label{fig:verysmallg1}
\end{figure}

\begin{figure}[h]
	\begin{subfigure}{.5\textwidth}
		\centering
		\includegraphics[width = 7.0cm]{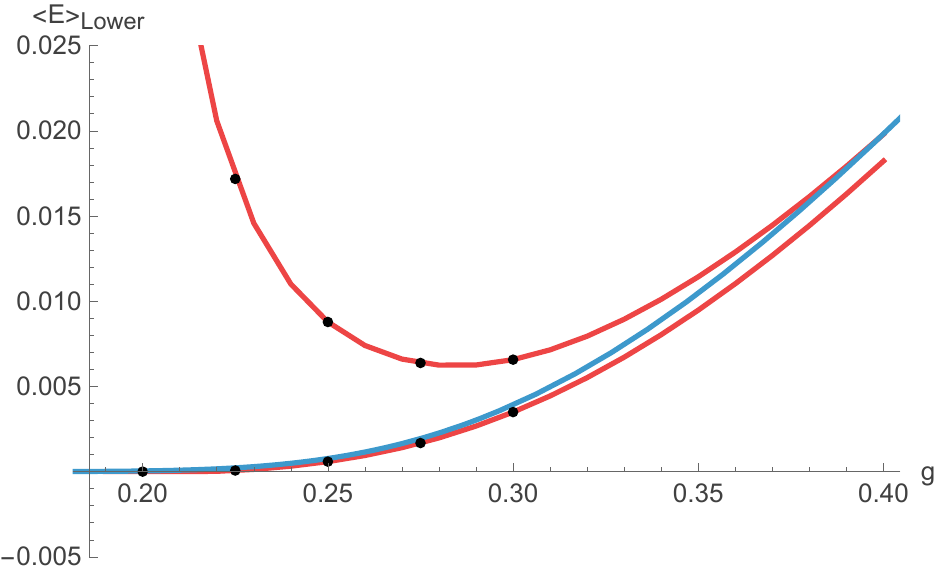}
	\end{subfigure}
	\begin{subfigure}{.5\textwidth}
		\centering
		\includegraphics[width = 7.0cm]{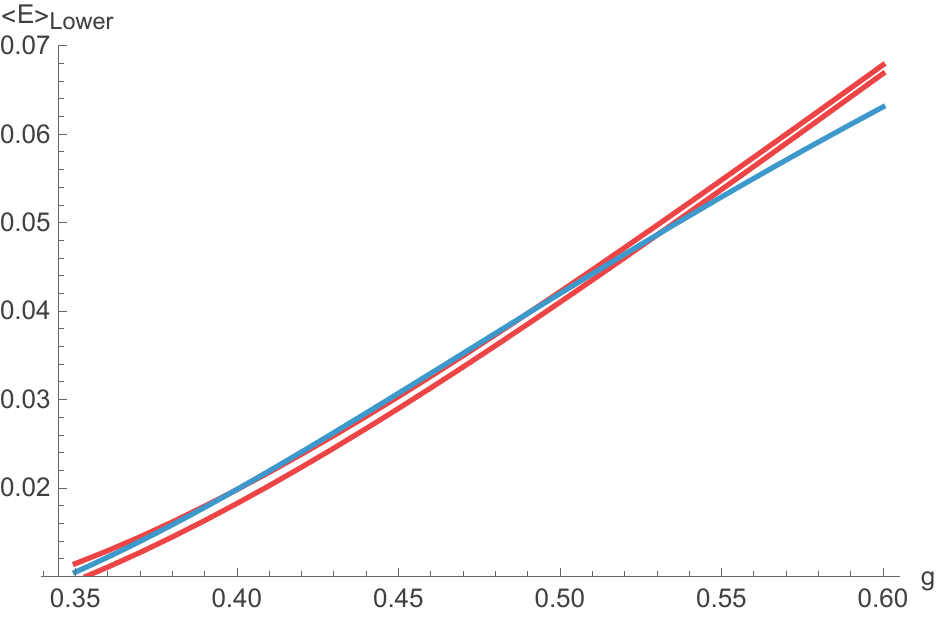}
	\end{subfigure}
    \caption{Ground state energy for SUSY QM with $W = 1/2 x^2 + g/3 x^3$ using level 10 constraints. The red line was made with mathematica's solver and the black dots are SDPB's result. The blue line is $E_0$ from equation~\eqref{eq:instanton}.}
    \label{fig:verysmallg2}
\end{figure}

The results of the small $g$ region are shown in figure~\ref{fig:smallgn1}. We see that they converge fairly well except at very small $g$. For very small $g$, we have computed the bounds using a combination of SDPB and mathematica. Figure~\ref{fig:verysmallg1} shows the small $g$ bounds obtained using SDPB. The plot is sparse because each point takes about 10 minutes to run. For larger $g$, we have checked where mathematica and SDPB start to agree and then plotted many more points using Mathematica's solver in figure~\ref{fig:verysmallg2}. The minimum of the energy at small coupling is due to SUSY-breaking instantons. These can be shown to contribute to the energy \cite{Tong} as
\begin{align}
\label{eq:instanton}
    E_0= \frac{1}{2 \pi} e^{-2 S_\text{inst}} \, \qquad S_\text{inst}  = |W(x_1) - W(x_2)| = \frac{1}{6 g^2}
\end{align}
where $x_1 = 0$ and $x_2 = -1/g$ are the solutions to $W'(x) = 0$. We have included this line in figure~\ref{fig:verysmallg2} in blue: we see that it is a good approximation to the allowed region until sometime before $g = .4$, when the $1 / S_\text{inst}$ corrections to these formulae become large.

\section{Supersymmetric matrix quantum mechanics}

Now we turn to supersymmetric matrix quantum mechanics. Like for the single SUSY particle on a line, there exists a whole class of SUSY matrix theories, specified by a general superpotential $W$. Such theories were introduced by Marinari and Parisi \cite{Marinari:1990jc}, who primarily considered the cubic superpotential. For a general superpotential the action is\footnote{This is the action of the so-called ``complex formulation'' -- it is related to the real formulation by 
\begin{align}
    \Psi_1 = \frac{1}{\sqrt{2}} \left( \Psi + \Psi^\dagger\right) \, ,\qquad
\Psi_2 = \frac{1}{\sqrt{2}i} \left( \Psi - \Psi^\dagger\right) \, .
\end{align}}
\begin{align}
    S = \frac{1}{2} \int dt   \left(  \dot{X}_{ij} \dot{X}_{ji}  - \frac{\partial W(X)}{\partial X_{ij}} \frac{\partial W(X)}{\partial X_{ji}} - i \left( \Psi_{ij}^\dagger \dot{\Psi}_{ji} + \Psi_{ij} \dot{\Psi}_{ji}^\dagger \right) -  [\Psi^\dagger_{ij} , \Psi_{kl}] \frac{\partial^2 W(X)}{\partial X_{ij} \partial X_{kl}}\right) \, .
\end{align}
We define the momenta
\begin{align}
   P_{ij} \equiv \Pi_{X_{ji}}  =  \dot{X}_{ij} \, , \qquad 
\Pi_{\Psi_{ij}} = - i  \Psi^\dagger_{ji} \, , 
\end{align}
This means that $P_{ij}$ is conjugate to $X_{ji}$ and $\Psi^\dagger_{ij}$ to $\Psi_{ji}$, leading to the commutation relations:
\begin{align}
    [X_{ij} , P_{kl}] = i \delta_{il} \delta_{jk} \, , \qquad 
\{ \Psi^\dagger_{ij} , \Psi_{kl} \} = \delta_{il} \delta_{jk} \, ,
\end{align}
with the other commutators / anticommutators being $0$, and giving the Hamiltonian:
\be
\label{eq:MPHam}
H =  \frac{1}{2} P_{ij} P_{ji} + \frac{1}{2} \frac{\partial W(X)}{\partial X_{ij}} \frac{\partial W(X)}{\partial X_{ji}} + \frac{1}{2}  [\Psi^\dagger_{ij} , \Psi_{kl}] \frac{\partial^2 W(X)}{\partial X_{ij} \partial X_{kl}}  \, .
\ee
The system described by equation~\eqref{eq:MPHam} has the following symmetries:

\paragraph{Supersymmetry}

The model is invariant under the SUSY transformations  
\begin{align}
    \delta X_{ij} \ &= \ i \xi_1 \Psi_{ij} +  i \xi_2 \Psi_{ij}^\dagger \, ,\\
    \delta \Psi_{ij} \ &= \ \xi_2 \dot X_{ij} + i \xi_2 \frac{\partial W(X)}{\partial X_{ji}} \, , \\
    \delta \Psi_{ij}^\dagger  \ &= \ \xi_1 \dot X_{ij} - i \xi_1 \frac{\partial W(X)}{\partial X_{ji}} \, ,
\end{align}
The supersymmetry transformations have associated supercharges
\begin{align}
    \label{eq:superchargesaa}
    Q \ &= \  \left( P_{ij} + i \frac{d W}{d X_{ji}} \right) \Psi_{ji} \, , \qquad 
    \bar Q \ = \ \left( P_{ij} - i \frac{d W}{d X_{ji}} \right) \Psi_{ji}^\dagger \, .
\end{align}
One can check that these supercharges satisfy the equations
\begin{align}
    \{ Q, \bar Q \} = 2H \, , \qquad \{ Q, Q \} = \{ \bar Q, \bar Q \} = 0 \,.
\end{align}

\paragraph{Gauge symmetry} Invariance under the $SU(N)$ gauge symmetry requires that the MP action is symmetric under
\begin{align}
    X \to U(t)^\dagger X U(t)  \, , \qquad \Psi \to U(t)^\dagger \Psi  U(t) \, , \qquad \Psi^\dagger \to U(t)^\dagger \Psi^\dagger U(t) \, .
\end{align}
This implies the constraint 
\begin{align}
    \left< \tr G \mathcal O \right> \ = \ 0  
\end{align}
for any operator $\mathcal O$. Here $G$ is the traceless gauge operator defined by
\begin{align}
    \label{eq:gauge}
    G = i [X, P] - (\Psi \Psi^\dagger + \Psi^\dagger \Psi ) + 2  N I \, .
\end{align}

\paragraph{Fermion number symmetry} Just as in the $N = 1$ case, the fermion number $F
= \tr ( \Psi^\dagger \Psi) $ is an additional charge which commutes with the Hamiltonian. It takes values between
\begin{align}
    0 \leq F \leq N^2 \, .
\end{align}


\paragraph{Positivity}

We have positivity in the form
\begin{align}
    \M \succeq 0 \, ,
\end{align}
where $\M_{ij} = \langle O^\dagger_i O_j \rangle$. We shall choose a small basis of $\mathcal{O}_i$ formed from strings of $X$s, $P$s, and $\Psi^\dagger \Psi$.

The expectation values comprising the matrix $\M$ are subject to further relations deriving from the Heisenberg equation
\begin{align}
    \langle [H, \mathcal{O}] \rangle = 0 \,,
\end{align}
the gauge constraint,
\begin{align}
    \langle G \mathcal{O} \rangle = 0 \, .
\end{align}
and Fermion number symmetry
\begin{align}
    \langle [F, \mathcal{O}] \rangle = 0\,.
\end{align}

\paragraph{Supercharge constraints}

Another constraint satisfied by SUSY systems is 
\begin{align}
    \langle [Q^\dagger Q , \mathcal O] \rangle = 0 \, .
\end{align}
This is because $Q^\dagger Q$ commutes with the Hamiltonian, and so the energy eigenstates can be chosen to be in a basis of definite $Q^\dagger Q$. We found in practice that these constraints are satisfied but they are redundant -- they are automatically implied by the constraints above, at least at the level we are working.

\paragraph{Ground state constraint}

A final constraint that we impose is 
\begin{align}
    \langle \mathcal O_i^\dagger [H, \mathcal O_j] \rangle \succeq 0 \, .
\end{align}
This is a constraint that applies only to the ground state, which will be our primary interest. It arises as the zero temperature limit of the thermal states discussed in \cite{Fawzi:2023fpg, Cho:2024kxn}.

\subsection{Quadratic superpotential}

The quadratic superpotential will be a useful warmup to check our bounds and give some examples of the constraints that arise. We take
\begin{align}
    W(X) = \frac{1}{2} a \Tr X^2 \,.
\end{align}%
The result is a Hamiltonian where the fermions and bosons are decoupled,
\begin{align}
    H = \frac{1}{2} \tr \left( P^2 +  a^2  X^2 +  a [\Psi^\dagger, \Psi] \right) \, .
\end{align}
formed from supercharge $Q = \tr (P + i a X) \Psi$. Here $a$ plays the role of a mass, and we see that the system is only supersymmetric if the mass in the fermionic sector and bosonic sector is the same. We can uncover a few non-trivial bounds using small bootstrap matrices. For instance, choosing $ \{ I, \, X, \, P \}$ as a basis and deriving the constraints $\tr XP = -\tr PX = i N^2 / 2$, $\tr P^2 = a^2 \tr X^2$, we find (suppressing the necessary expectation values)
\begin{align}
    \begin{pmatrix}
        \tr I  & \tr X & \tr P \\
        \tr X & \tr X^2 & \tr XP \\
        \tr P & \tr PX & \tr P^2 
    \end{pmatrix} \ = \ 
    \begin{pmatrix}
        N  & 0 & 0 \\
        0 & \tr X^2 & \frac{i N^2}{2} \\
        0 & -\frac{i N^2}{2} & a^2 \, \tr X^2
    \end{pmatrix} \ \succeq 0 \, .
\end{align}
Assuming without loss of generality that $a > 0$, this implies 
\begin{align}
     \left< \tr X^2 \right>  \geq \frac{N^2}{2 a} \,.
\end{align}
A simple check for these constraints and this bound is that it gives us, for the bosonic part of the Hamiltonian
\begin{align}
   \langle H_B \rangle  \ = \ \left< \frac{1}{2} \tr P^2 + \frac{a^2}{2} \tr X^2 \right> \ = \ a^2 \left< \tr X^2 \right> \ \geq \ a \frac{N^2}{2} \, .
\end{align}
The right hand side is the ground state energy of the (matrix) quantum harmonic oscillator, whose Hamiltonian is $H_B$. The constraints did not require the fermionic part in their derivation so they are valid also for the SHO, and we see that the simple lower bound on the energy is saturated by the true ground state. 

Forming another bootstrap matrix from the basis $\Psi$, $\Psi^\dagger$, we find
\begin{align}
    \begin{pmatrix}
        \tr \Psi^\dagger \Psi & \tr \Psi^\dagger \Psi^\dagger \\
        \tr \Psi \Psi & \tr \Psi \Psi^\dagger 
    \end{pmatrix} \ \succeq 0 \, .
\end{align}
This tells us that the fermion number $F = \tr \Psi^\dagger \Psi$ is positive. Furthermore since $\tr \Psi \Psi^\dagger = N^2 - F$ from the commutation relations, we also have that $N^2 - F > 0$. So we find 
\begin{align}
    0 \leq F \leq N^2 \, .
\end{align}
Using all of our constraints, we can write the expectation of the Hamiltonian as 
\begin{align}
    \langle H \rangle = \left<  a^2 \tr X^2 + \frac{a}{2} (2F - N^2) \right> > 0 \, .
\end{align}
where the right-hand side (zero) is obtained by considering the minimum possible value for the bosonic and fermionic parts separately. We can easily see from the bootstrap that the energy cannot be negative, and that if there is a zero-energy state, it must have $F = 0$. Note that if we had assumed $a<0$ from the beginning, then we would have found $F = N^2$ to be the ground state instead.

\paragraph{Numerical bounds}

\begin{figure}[h]
	\begin{subfigure}{.5\textwidth}
		\centering
		\includegraphics[width = 7.0cm]{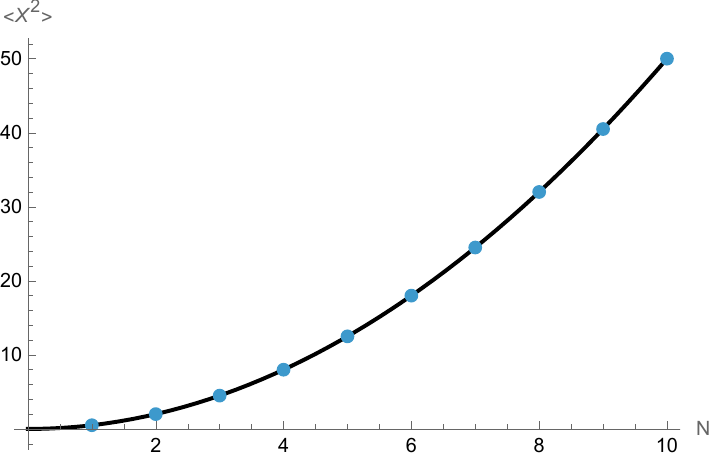}
	\end{subfigure}
	\begin{subfigure}{.5\textwidth}
		\centering
		\includegraphics[width = 7.0cm]{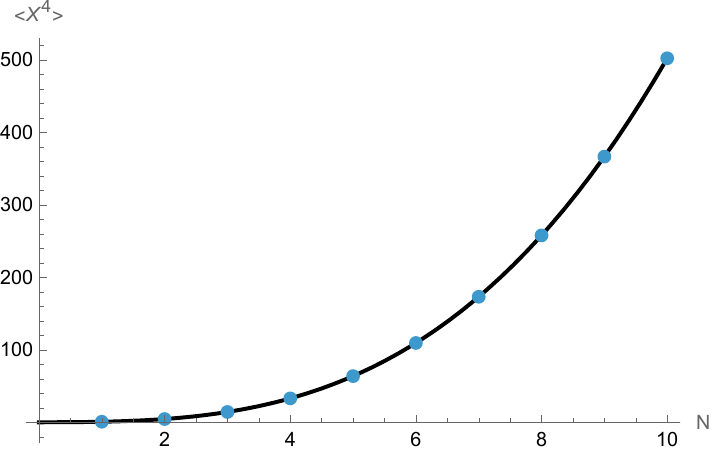}
	\end{subfigure}
    \caption{Ground state lower bounds for expectations in the quadratic superpotential with $a = 1$. We took a $7 \times 7$ matrix generated from the set of operators given in ~\eqref{eq:quadops}. The exact results are represented by black line, while the bootstrap bound is represented in blue.}
    \label{fig:matrixquad1}
\end{figure}

Tight bounds can be obtained by considering larger systems. In fact, we found that nearly exact values for the ground state can be obtained at finite system size. Using the set of operators 
\begin{align}
    \label{eq:quadops}
    \{ I, X, P, \Psi, \Psi^\dagger, X^2, PX, P^2, \Psi \Psi^\dagger , \Psi X, \Psi P, \Psi^\dagger X, \Psi^\dagger P\} \, ,
\end{align}
we found the lower bounds displayed in figure~\ref{fig:matrixquad1}.

\subsection{Cubic superpotential}

Now we will move to the cubic superpotential. The original MP model studied in \cite{Marinari:1990jc} involves a quadratic and cubic term, so that is what we will study here. With the superpotential 
\begin{align}
    W = \frac{1}{2} X^2 + g \frac{1}{3} X^3, 
\end{align}
the Hamiltonian becomes 
\begin{align}
    H \ &= \ 
    \frac{1}{2} \, \tr \left(  P^2 +  X^2 + 2 g X^3  + g^2 X^4  + [ \Psi^\dagger, \Psi] + 
    g  ([\Psi^\dagger, \Psi] X + X [\Psi^\dagger, \Psi]  )  \right) \, .
\end{align}
This theory has supercharges
\begin{align}
    \label{eq:superchargesbb}
    Q \ &= \  \left( P_{ij} + i X_{ij} + i g X_{ik} X_{kj} \right) \Psi_{ji} \, , \qquad 
    \bar Q \ = \ \left( P_{ij} -  i X_{ij} - i g X_{ik} X_{kj}  \right) \Psi_{ji}^\dagger \, .
\end{align}
Like any model with where the leading power of $X$ in the superpotential is odd, this theory does not have a normalizable zero-energy state. That means that by imposing $\langle 1 \rangle = N$, we are singling out normalizable states and thus can expect a positive bound for the energy. 

\begin{figure}[h]
		\centering
		\includegraphics[width = 8.0cm]{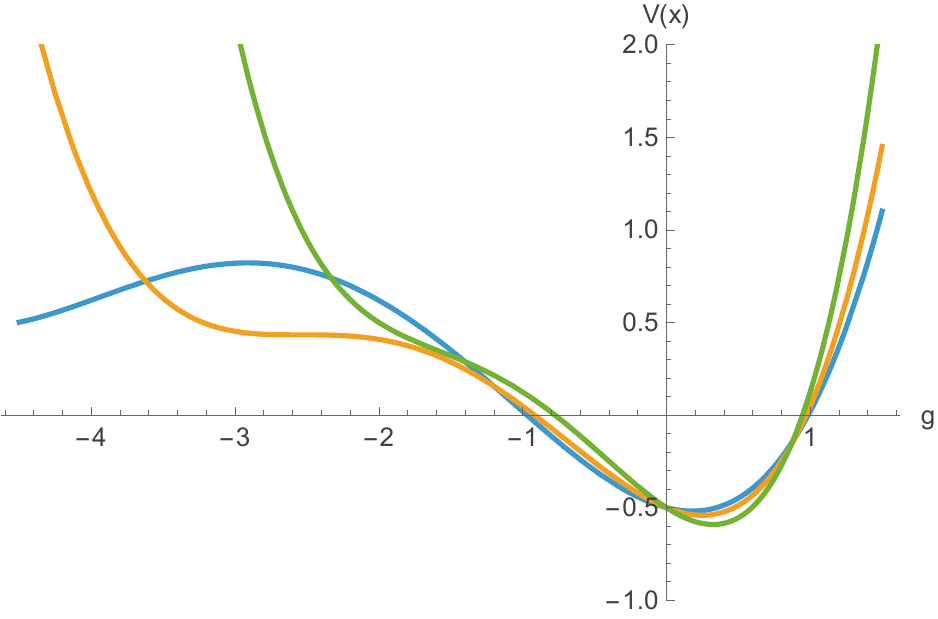}
    \caption{Potential of the cubic matrix model in the bosonic sector: blue is $g = .2$, orange is $g = g_c$, and green is $g = .5$. One sees that for $g = g_c$, the upper extrema merge to an inflection point. }
    \label{fig:potential}
\end{figure}

For $N = 1$ in the bosonic sector, the system has potential
\begin{align}
    V(x) = \frac{1}{2} \left( x^2 + 2 g x^2 + g^2 x^4 - 1 - 2  g x \right) \, . 
\end{align}
At large $N$, the bosonic sector is described by $N$ non-interacting fermions with the same potential. $V$ has a critical point 
at $g_c^2 = 1 / (6 \sqrt{3}) $ -- below $g_c$ the potential has two minima, and above it there is only one. We have displayed this in figure~\ref{fig:potential}

Below the critical point, the energy vanishes in perturbation theory, but instantons will give non-perturbative corrections to the energy.

\subsubsection{Bootstrap}

In the language of \cite{Lin:2024vvg}, where the level of $X$ is 1, $P$ is 2, and $\Psi$ and $\Psi^\dagger$ are each 3/2, we have performed a level-8 bootstrap of this system. This means that we have formed our matrix from all strings that are level 4 or lower, and the bootstrap matrix contains every operator of level 8 or lower. There are 44 such strings, leading to a $44 \times 44$ bootstrap matrix.

Fermion number conservation leads to the constraint that any operator with a different number of $\Psi$s and $\Psi^\dagger$s is equal to zero. This is because $\Psi^\dagger$ and $\Psi$ are raising / lowering operators for fermion number so they map states to orthogonal sectors. The resulting constraint is useful for removing a large number of expectation values. The bootstrap matrix becomes block diagonal: we can separate the matrices formed form fermion-number-0 strings ($1$, $X$, $P$, $\Psi \Psi^\dagger$, ...) and fermion-number-1 strings ($\Psi$, $\Psi X$, $\Psi P$, $\Psi \Psi \Psi^\dagger$, ...) and so on. Up to level 8, this leads to 5 bootstrap matrices: 
\begin{align}
    \text{fermion number} = 0 \qquad & 20 \times 20 \nonumber \\
    \text{fermion number} = \pm 1 \qquad & 8\times 8 \nonumber \\
    \text{fermion number} = \pm 2 \qquad & 4 \times 4 \nonumber
\end{align}
These matrices have a total 143 free variables before any constraints have been used, though fermion number is manifest in this formulation, and we have already used the algebra to put all traces in a canonical form. 

The relevant constraints, in the order we imposed them, are (1) reality, (2) gauge symmetry, and (3) equations of motion. The reality constraints say that
\begin{align}
    \langle \mathcal O_1 ... \mathcal O_n \rangle^* = \langle \mathcal O_n^\dagger ... \mathcal O_1^\dagger \rangle \, .
\end{align}
$X$ and $P$ are Hermitian, while $\Psi$ and $\Psi^\dagger$ transform into each other. One can choose real wavefunctions, meaning that $\langle \mathcal O_1 ... \mathcal O_n \rangle^* = \pm\langle \mathcal O_1 ... \mathcal O_n \rangle$, with the minus being chosen when there are an odd number of $P$s among the $\mathcal O_i$. In total, we found that the reality constraints reduce the total number of variables to 9. The gauge constraints are defined in ~\eqref{eq:gauge}. This reduces the total number of variables to 44.

In both the gauge and reality constraints, we found that we needed to go to rather high levels before we stopped getting ``useful'' constraints (those involving only operators of level 8 and lower). In practice, we continued to find useful constraints even at length 11 for reality and length 8 for gauge (the gauge operator itself is level 3). 

Finally the equations of motion take the form $ \langle [H, \mathcal O] \rangle = 0$. The operator $ [H, \mathcal O]$ has level at most one higher than the level of $\mathcal O$. So we impose the equations of motion on all operators of level 7 or above. All said and done, our SDP has 27 free variables

In addition to the above constraints, we also used the ``ground state constraints'' that arise from considering the thermal bootstrap \cite{Fawzi:2023fpg, Cho:2024kxn} in the zero-temperature limit. These constraints amount to the positive-definiteness constraint
\begin{align}
    \langle \mathcal O_i^\dagger [ H , \mathcal O_j] \rangle \succeq 0
\end{align}
These constraints become high-level fairly quickly, so they only added a 5x5 matrix to the positivity conditions. In our case they did not change the bounds much and they were not enough to obtain upper bounds on the energy, but it did make the SDP solver converge faster. 

\subsubsection{Results}

Semidefinite programming allows us to get bounds on various quantities in the ground state. The data for some of the plots was generated using Mathematica's SDP solver. In the small$-g$ region, Mathematica's solver was not trustworthy and we used SDPB \cite{Simmons-Duffin:2015qma}. A further note: all of our bounds were made by rescaling $g \to g / \sqrt{N}$ to facilitate easier comparison with \cite{Marinari:1990jc}. 

\begin{figure}[h]
	\begin{subfigure}{.48\textwidth}
		\centering
		\includegraphics[width = 7cm]{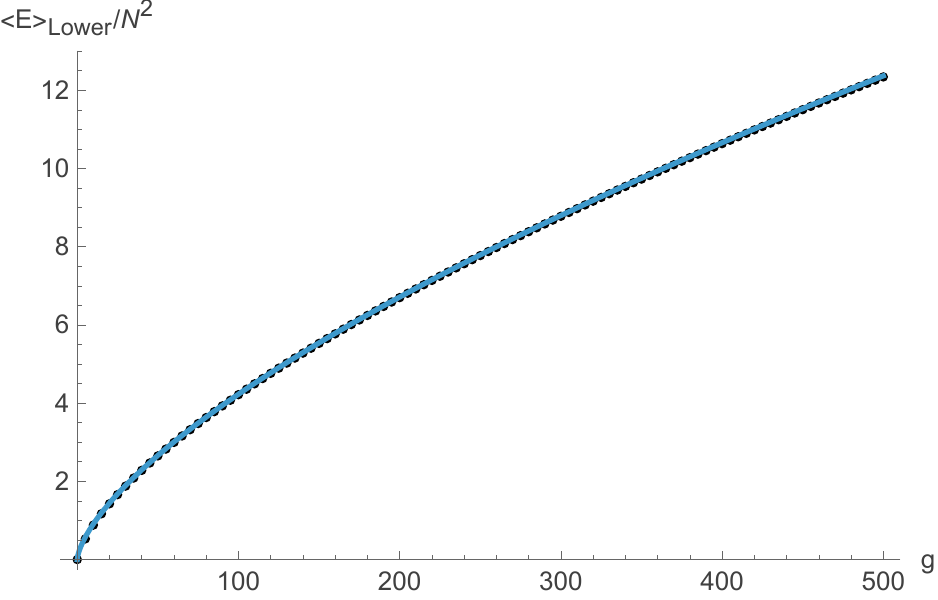}
	\end{subfigure}
	\begin{subfigure}{.48\textwidth}
		\centering
		\includegraphics[width = 7cm]{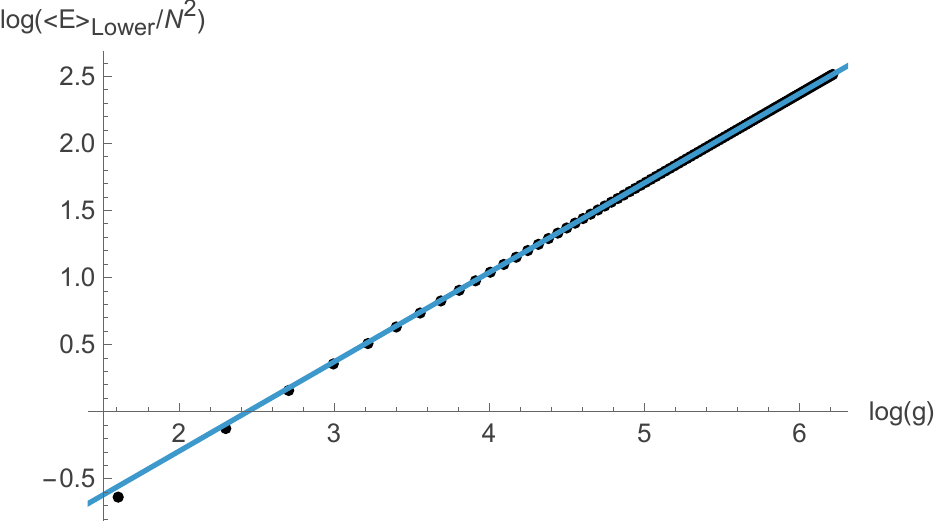}
	\end{subfigure}
    \caption{Energy lower bounds over $N^2$ vs. $g$ at large $g$, $N = 100$. The blue curve represents the fit $.196 \, g^{2/3}$.}
    \label{fig:largeg}
\end{figure}

\paragraph{Large $N$ and $g$} We have displayed the bounds on the energy at large $N$ and large $g$ in figure~\ref{fig:largeg}. We find that the lower bound scales as $g^{2/3}$. It also scales as $N^2$ -- figure~\ref{fig:largeg} was made for $N = 100$, but the $N = 1000$ plots are nearly identical. The blue line was created by fitting the linear part of the log-log plot -- the slope of that line was found to be $.667$ and the $y$-intercept was about $-1.63$.

Like for $N = 1$ in section~\ref{sec:qm_bootstrap}, this scaling is evident by rescaling $X \to g^{-1/3} X $ and $P \to g^{1/3} P$. This leads to the Hamiltonian
\begin{align}
    H = \frac{1}{2}g^{2/3} \left( P^2 + X^4 + ([\Psi^\dagger, \Psi] X + X [\Psi^\dagger, \Psi]  ) + \mathcal O(g^{-2/3}) \right)
\end{align}
so at large $g$, the energy is proportional to $g^{2/3}$, with some corrections at higher orders of $g^{-2/3}$. This may be modeled as 
\begin{align}
    \frac{E_0}{N^2} = \kappa_0 g^{2/3} + \kappa_1 + \kappa_2 g^{-2/3} + ...
\end{align}
We find $\kappa_0 \simeq .196$. 

We can estimate $\kappa_0$ in the bosonic sector using the WKB approximation. The system is equivalent to $N$ free fermions each with the potential $V$. The Fermi energy $\varepsilon_F$ is given by the condition
\begin{align}
    N \ = \ \frac{1}{\pi} \int dx \sqrt{2(\varepsilon_F - V(x))} 
\end{align}
which gives $\varepsilon \simeq 1.006 \, N^{4/3}$. The total energy is given by 
\begin{align}
    E \ = \ \frac{ \sqrt{2}}{3\pi} \int dx \left((\varepsilon_F - V(x))^{3/2}+ 3 V(x) \sqrt{\varepsilon_F - V(x)}  \right) \ = \  .242 \,  g^{2/3} N^{7/3}
\end{align}
which, after rescaling $g \to g / \sqrt{N}$, gives $\kappa_0 = .242$. We believe that the ground state should be purely bosonic, as fermions have double spaced energies (see \cite{McGreevy:2003dn} for a discussion). Adding fermions also decreases the coefficient of the linear term in the potential, which increases the energy. So we do not know what the source of the discrepancy is -- the most likely case is that it is a truncation error and will increase as we increase the size of the basis (though there was no increase from level 7 to 8, as we discuss below).

\paragraph{Small $g$}

The other particularly interesting regime is near the critical point, $g_c = (6 \sqrt{3})^{-1/2}$. At large $N$ in the particular double-scaling limit where $N (g - g_c)^{5/2}$ is held fixed, graphs of all topologies contribute and the theory is expected to be described by a supersymmetric string theory. A candidate, the worldline theory of unstable $D0$ branes, was described in \cite{McGreevy:2003dn}. 

\begin{figure}[h]
	\begin{subfigure}{.32\textwidth}
		\centering
		\includegraphics[width = 4.8cm]{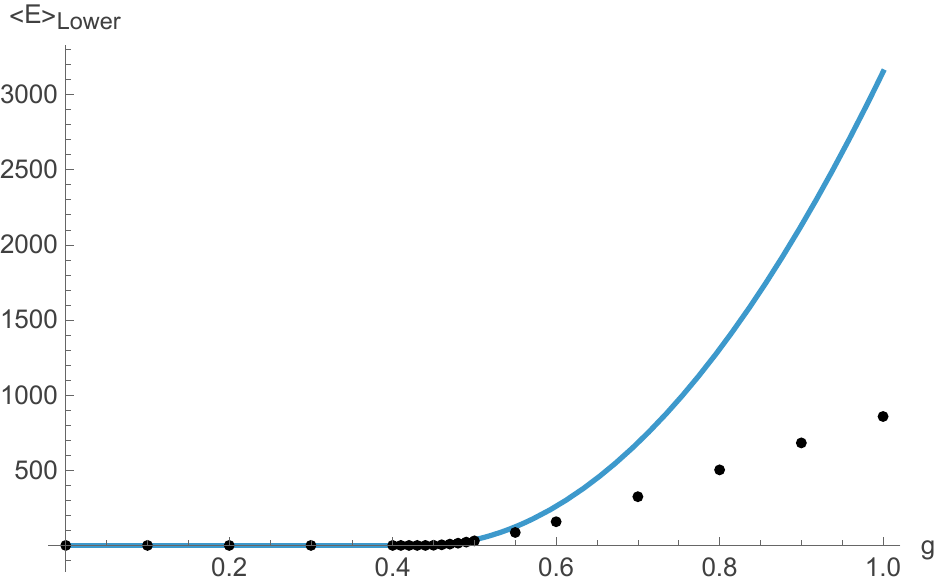}
	\end{subfigure}
	\begin{subfigure}{.32\textwidth}
		\centering
		\includegraphics[width = 4.8cm]{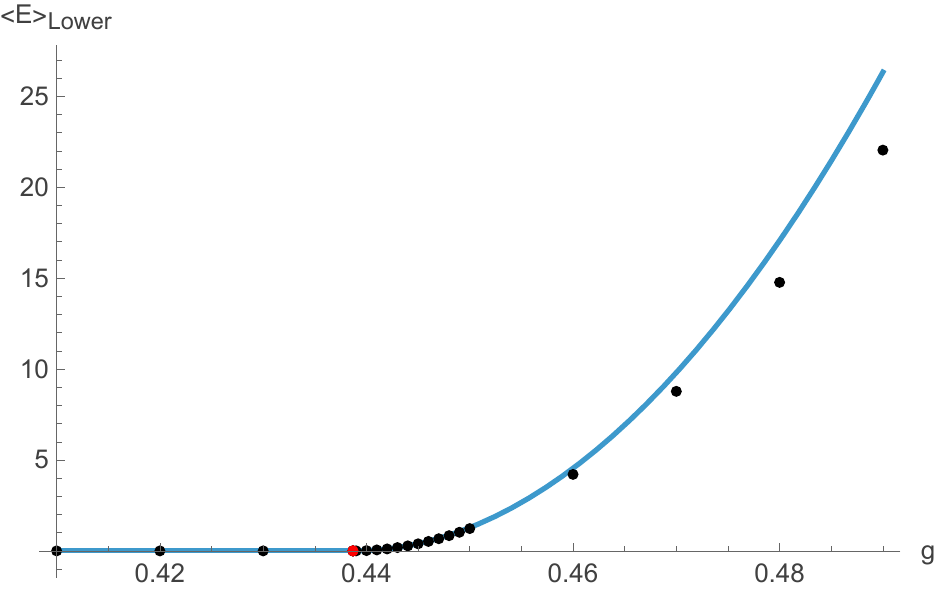}
	\end{subfigure}
    \begin{subfigure}{.32\textwidth}
		\centering
		\includegraphics[width = 4.8cm]{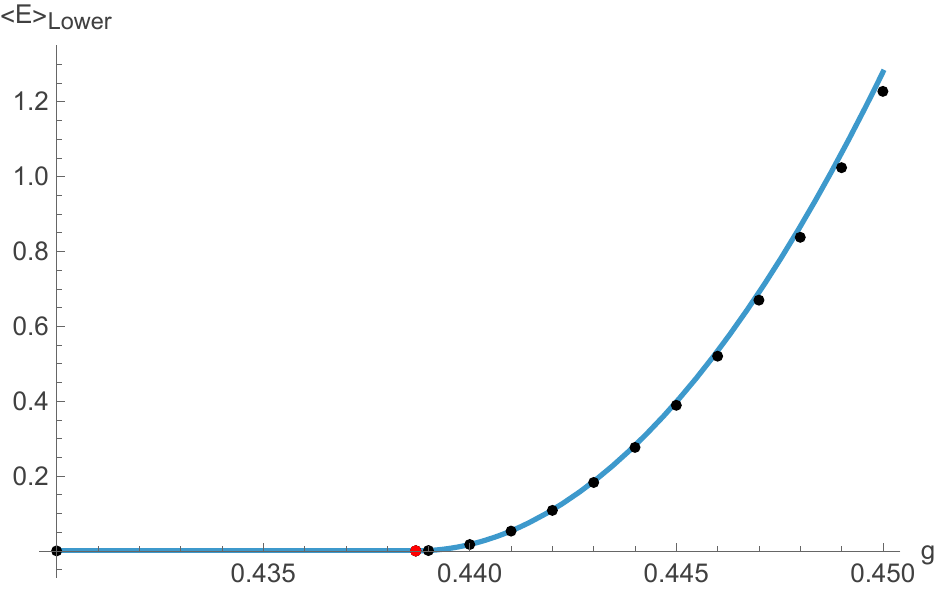}
	\end{subfigure}
    \caption{Energy lower bounds as a function of $g$. Black dots represent the numerical lower bound, and the red dot is the bound at $g_0$ -- all bounds were obtained with SDPB.}
    \label{fig:n100}
\end{figure}

\begin{figure}[h]
		\centering
		\includegraphics[width = 8.0cm]{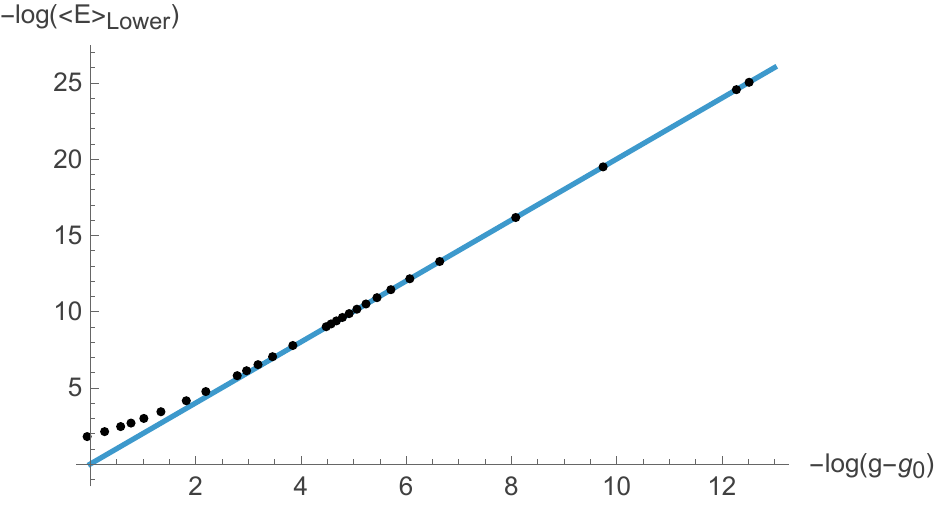}
    \caption{Log-log plot of energy vs. $g - g_c$, demonstrating the accuracy of the $E_\text{lower} = (g - g_0)^2$ scaling as $g_0$ is approached.}
    \label{fig:n100log}
\end{figure}

We have plotted our lower bounds near the critical point for $N = 100$\footnote{We have compared, for several points at small $g$, with the bounds at $n = 1000$. We find that the ratio of lower bounds is 100 to within $10^{-11}$, in agreement with the $N^2$ scaling of the energy at large $N$.}. Our results have the bizarre feature that they are nonanalytic at $g_0 = \sqrt{2} \, g_c$ (see figure~\ref{fig:n100}), and scale at the critical point as $\langle E \rangle_\text{Lower} = (g-g_0)^2$ (see figure~\ref{fig:n100log}). 

\begin{figure}[h]
		\centering
		\includegraphics[width = 8.0cm]{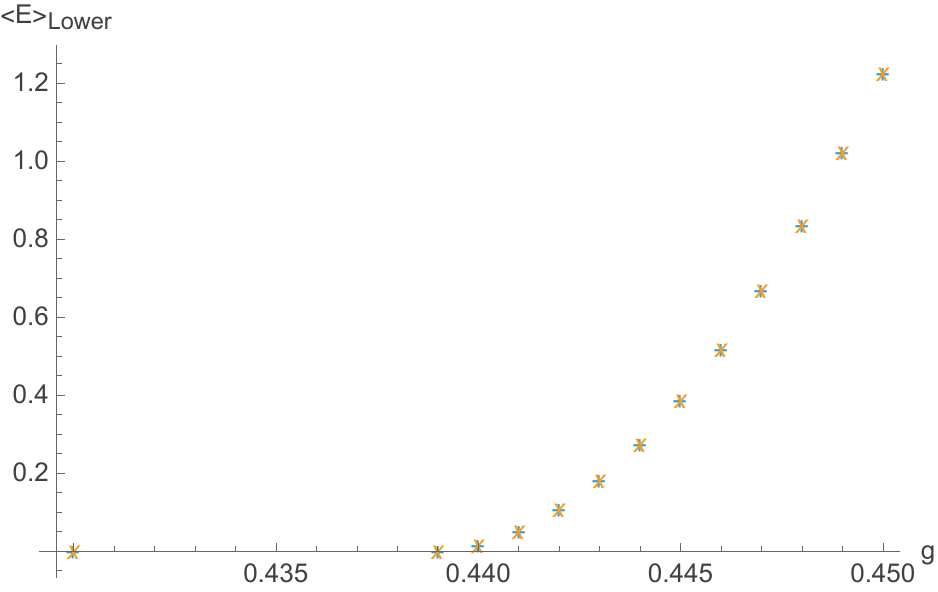}
    \caption{Comparison of the energy lower bounds for level 7 (blue ``$+$'') and level 8 (orange ``$\times$'') bootstrap matrices.}
    \label{fig:levelcomp}
\end{figure}

\paragraph{Convergence} In figure~\ref{fig:levelcomp}, we have compared the lower bounds near the critical point for the level 8 bootstrap with the level 7 results. The level 7 bootstrap matrix is $27 \times 27$, compared to the $44 \times 44$ level 8 matrix. SDPB runs roughly 10 times faster for this smaller matrix. We have also performed the bootstrap at level 6, but the results were extremely weak -- about $-2800$ for every point included in figure~\ref{fig:levelcomp}. We do not know why the bounds weaken so dramatically at level 6. One can see that the results at level 7 and 8 are essentially on top of each other -- in fact, the differences are each less than $10^{-9}$. However, at $N = 1$, we found cases where increasing the level had no effect on the bounds, while further increases led to large improvements. So we are not sure how well the bounds have converged. Understanding more will require a level-9 bootstrap or higher. However, our current results take about 8 hours to compute the constraints (by which we mean doing the algebra to generate the bootstrap matrices) on a laptop. Each point at level 8 takes 5-10 minutes to compute in SDPB. So a level 9 bootstrap is beyond what we can do with our current computational resources\footnote{We tried a ``partial level 9 bootstrap'' where we use operators up to level 9 but only derive a subset of the constraints. However we have not been able to get stronger bounds doing this so far.}.

\section{Discussion}

In this paper we have performed a numerical bootstrap of several supersymmetric models, with the aim of determining the most general set of constraints needed to bootstrap the MP model, a SUSY matrix model that conjecturally describes $D0$ brane dynamics. Using a non-linear solver for the $N = 1$ case, we found tight bounds on the allowed regions in the space of expectation values. For the quadratic superpotential we saw that there was a unique zero-energy ground state, with all other states paired, and for the cubic superpotential we found that the lowest energy state has positive energy and degeneracy two. This signals that the zero-energy ground state, which is non-normalizable in the cubic model, is excluded by the assumptions of the bootstrap --namely, that $\langle 1 \rangle = 1$. 

For the matrix model, which generalizes SUSY QM to arbitrary $N$, we found lower bounds on the ground state energy. For the quadratic model these bounds were tight even with a fairly small matrix. For the cubic model, we constructed a bootstrap matrix of all operators up to level 8. This allowed us to reproduce the scaling (but not leading coefficient) of the MP model predicted by the WKB approximation at large $N$ and large coupling $g$. An interesting feature at small $g$ is a sharp kink at almost exactly $\sqrt{2} \,  g_c$, where $g_c$ is the critical point where the model goes from a double-well to a single well.

The set of constraints we used included (1) reality of the wavefunctions, (2) gauge symmetry $\langle \tr G \mathcal O \rangle = 0$, and (3) the equations of motion $\langle [H, \mathcal O] \rangle = 0$. We also used fermion number conservation to set a number of expectation values to zero from the outset. We tested constraints specific to supersymmetry, which were $\langle [Q^\dagger Q, \mathcal O] \rangle = 0$. We found that these constraints, while true, turned out to be redundant with (1), (2), and (3). Perhaps at higher-levels these constraints will be necessary.

In principle the algorithm described here and familiar from other works could be pushed to arbitrarily high-level. From experience, the bootstrap bounds tend to converge fairly quickly as the level is increased. However, we have seen many of the broad features of the model so it is not clear what precise numbers would be particularly useful to try to bootstrap. One potentially interesting target is the action of the instanton which breaks SUSY in the low-coupling phase of the MP model. The energy at $g < g_c$ is exactly zero in perturbation theory, and our numerical bounds were of the order of $10^{-21}$, which is essentially zero for our purposes. However, there are instantons representing tunneling between the two wells whose contributions to the energy is expected to be %
\begin{align}
    E_0 \simeq e^{-k s^2} 
\end{align}
where $s = (g - g_c) N^{4/5}$ and $k$ is an order 1 constant. For $N = 1$, our results were precise enough to match against these, and it would be interesting to see if this number could be determined precisely at large-$N$. Perhaps relatedly, increasing the level might allow for a precise upper bound on the ground state energy ground state constraints from the thermal bootstrap \cite{Fawzi:2023fpg, Cho:2024kxn}. For our level-8 bootstrap, these constraints comprised a very small matrix and had a limit effect. In particular, no upper bound could be found. 

Another interesting direction would be to use the bootstrap to compare the gauged vs. ungauged model. In the BFSS model, the difference between the gauged and ungauged theories was analyzed in \cite{Maldacena:2018vsr}, where it was shown that the non-singlet operators present in the ungauged model are very heavy, leading the theories to be similar at low-energies. This is interesting in light of our results, where gauge constraints were required to remove an enormous number of free variables: in fact, we were unable to solve the SDP in the ungauged model. It would be very interesting to perform a higher-powered numerical comparison of the two theories to see the effect of gauging on the bootstrap bounds. 

By determining the constraints necessary to bootstrap the Marinari-Parisi model, we have paved the way to bootstrapping other similar systems. One good example would be the MP model with a logarithmic superpotential. In quantum mechanics, such a term leads to an inverse square potential, which is a marginal coupling in 0+1 dimensions. The resulting theory is known as conformal quantum mechanics \cite{Jackiw:1972cb, deAlfaro:1976vlx} and exhibits a rich set of features depending on the coupling (see e.g. \cite{Gitman:2009era}). In matrix quantum mechanics, the situation is much less well-understood. After integrating out the diagonalizing unitary transformations, the eigenvalues of the log potential form a consistent sub-sector described by the Calogero-Moser model \cite{Dabholkar:1991te, Rodrigues:1992by}, which has an inverse square potential between each eigenvalue. In \cite{Verlinde:2004gt}, it was proposed that $W = q \log X$ matrix model is dual to type II string theory on AdS${}_2$ that appears in the near-horizon limit of an extremal 2d black hole with flux $q$. It would be interesting to understand which observables can be matched between the matrix model side and the string side, so that the bootstrap could be used to probe the duality. The matrix model of \cite{Strominger:2003tm}, conjecturally dual to type 0A string theory on AdS${}_2$, would be a simpler test case, albeit one without supersymmetry. 

\clearpage

\section*{Acknowledgments}

We would like to thank Robert Brandenburger, Scott Lawrence, and Alex Maloney for useful discussions and correspondences. B.M. is supported by the Gloria and Joshua Goldberg Fellowship at Syracuse University and NSERC (Canada), with partial funding from the
Mathematical Physics Laboratory of the CRM. The part of S.L.'s work carried out at McGill University was supported by funding from the Fonds de recherche du Québec (FRQNT), NSERC, and the Canada Research Chair program.

\appendix
\addtocontents{toc}{\protect\setcounter{tocdepth}{1}}

\section{Constraining quantum mechanical systems using a convex semi-definite programming problem}
\label{ap:convex}

In the present section, we describe the method that allowed us to express $\langle H \rangle$ and $\mathcal{M}$ in terms of $\langle x^p \rangle$ in Section \ref{sec:quartic_corrections}, so that a minimum of $\langle H \rangle$ can be found by solving a convex semi-definite programming problem. 

For the sake of generality, let us consider the general system with the Hamiltonian
\begin{equation}
    H = \frac{1}{2} p^2 + V(x) \, .
\end{equation}
To express $\langle H \rangle$ in terms of $\langle x^p \rangle$, we make use of the Hamiltonian constraint $\langle [ H , p x ] \rangle = 0$ to obtain
\begin{equation}
\langle p^{2} \rangle =  \langle V'(x) x \rangle \, .
\end{equation}
Substituting $\langle p^{2} \rangle$ in $\langle H \rangle$ then yields
\begin{equation}
\langle H \rangle =  \frac{1}{2} \langle V'(x) x \rangle + \langle V(x) \rangle\, .
\end{equation}
In the present paper, we were interested in systems where $V(x)$ is a polynomial. For these potentials, the result above gives us an expression that only depends on $\langle x^p \rangle$. To express $\mathcal{M}$ in terms of $\langle x^p \rangle$, we first use relate the matrix elements of the block $C$ to the matrix elements of the block $B$ using the identity
\begin{equation}
    \langle x^n p^m \rangle = \sum_{k = 0}^{\text{min($n$,$m$)}}{ n \choose k} { m \choose k} k! i^k \langle p^{m-k} x^{n-k} \rangle ,
\end{equation}
which can be found by repeatedly using the commutation relations $[x,p] = i$ on $\langle x^i p^j \rangle$. We then recursively use the recursion relation
\begin{equation}
\langle p^{n} x^{m} \rangle= - \frac{m}{2} x^{m-1} - \frac{1}{m+1}\sum_{k=1}^{n-1} { n-1 \choose k} p^{n-k-1} i^{k+1} \langle V^{(k)}(x) x^{m+1} \rangle \, ,
\label{eq:general_O}
\end{equation}
on the remaining terms in $\mathcal{M}$ until only terms depending on $\langle x^p \rangle$ remain. Here, Equation \ref{eq:general_O} is found from the constraint $\langle [ H , O ] \rangle = 0$ using $\mathcal{O} = p^{i-1} x^{j+1}$ for $i \geq 1$ and $j \geq 0$.

When stopping at the step above and running a semi-definite programming algorithm, we found that the constraints described so far are not always sufficient to ensure convergence. In order to fix this issue, we also imposed the family of constraints
\begin{equation}
\sum_{k=1}^{n} { n \choose k} i^{k+1} p^{n-k} V^{(k)}(x) = 0 \, ,
\end{equation}
which can be found from $\langle [ H , O ] \rangle = 0$ when $O = p^{n}$.  When using Equation \ref{eq:general_O} recursively on the expression above, one finds an extra set of constraints that can be used to relate the $\langle x^p \rangle$'s among themselves. Imposing these extra constraints leads to a minimum of the energy that converges for the models we studied.

\section{Perturbation theory results for quartic corrections to the SUSY harmonic oscillator}
\label{ap:pert_theory}

In figure \ref{fig:Emin_vs_g_cubic_qm}, the bootstrap results are compared to perturbation theory estimates that can be found using textbook perturbation theory techniques. To get the estimates, we chose
\be
H = H_0 + \lambda V
\ee
such that 
\be
H_0 = \frac{1}{2} p^2 + \frac{1}{4g^2} + \frac{1}{2} x^2 \quad \text{and} \quad V = \frac{1}{4} g x^4 + \frac{1}{\sqrt{2}} \epsilon x \, ,
\ee
leaving $\lambda = g$ as a perturbation parameter. We then evaluated the perturbation theory estimate
\be
E_0 = E_0^{(0)} + \lambda \langle 0 | V | 0 \rangle + \lambda^2 \sum_{k \not= n} \frac{|\langle k | V | n \rangle|^2}{E_n^{(0)} - E_k^{(0)}} + ...
\ee
for the ground state energy at second order, where $E^{(0)}_0 = \frac{1}{2}+\frac{1}{4g^2}$ of the non-perturbed Hamiltonian $H_0$. Here, for simplicity of notation, we are taking $|n\rangle$ to be the eigenstates of the unperturbed Hamiltonian $H_0$. Keeping contributions of order $\mathcal{O}(g^2)$, we obtain
\be
E_0 = \frac{1}{2} + \frac{1}{4 g^2} + \frac{7 g^2}{16} + \mathcal{O}(g^3) \, ,
\ee
which is the desired result.

\section{ Symmetries of the Marinari-Parisi models}

Recall the action for a general superpotential is 
\begin{align}
    S = \frac{1}{2} \int dt   \left(  \dot{X}_{ij} \dot{X}_{ji}  - \frac{\partial W(X)}{\partial X_{ij}} \frac{\partial W(X)}{\partial X_{ji}} - i \left( \Psi_{ij}^\dagger \dot{\Psi}_{ji} + \Psi_{ij} \dot{\Psi}_{ji}^\dagger \right) -  [\Psi^\dagger_{ij} , \Psi_{kl}] \frac{\partial^2 W(X)}{\partial X_{ij} \partial X_{kl}}\right)
\end{align}
The momenta are defined by 
\begin{align}
    P_{ij} = \frac{\delta \mathcal{L}}{\delta \dot X_{ji}} = \dot X_{ij}\, , \qquad \Pi_{\Psi_{ij}} = \frac{\delta \mathcal{L}}{\delta \dot \Psi_{ji}} =  - i  \Psi^\dagger_{ij}
\end{align}
This leads to commutation relations 
\begin{align}
    [X_{ij}, P_{kl}] \ = \ i \delta_{il} \delta_{jk} \, , \qquad \left\{ \Psi^\dagger_{ij}, \Psi_{kl} \right\} = \delta_{il} \delta_{jk} \, .
\end{align}

\subsection{Gauge symmetry}

Invariance under the $SU(N)$ gauge symmetry requires that the MP action is symmetric under
\begin{align}
    X \to U(t)^\dagger X U(t)  \, , \qquad \Psi \to U(t)^\dagger \Psi  U(t) \, , \qquad \Psi^\dagger \to U(t)^\dagger \Psi^\dagger U(t)
\end{align}
We will consider infinitesimal transformations, where $U = 1 + i \Omega$, and then we will work to first order in $\Omega$. One the time-derivative parts give a non-zero contribution. One finds that 
\begin{align}
    \delta \left( \frac{1}{2} \dot X \dot X \right) = -i (  X \dot X -  \dot X X) \dot \Omega\\
    \delta \left( -\frac{i}{2} ( \Psi^\dagger \dot \Psi + \Psi \dot \Psi^\dagger)  \right)  =  (\Psi \Psi^\dagger + \Psi^\dagger \Psi ) \dot \Omega
\end{align}
Here $X$, $\Psi$, and $\Psi^\dagger$ are not operators but the fermions are still anticommuting variables, so cyclic permutations inside the trace may introduce minus signs. The total variation will be zero provided
\begin{align}
    \tr G_c \dot \Omega = 0 \, ,\qquad \text{for all } \dot \Omega \, .
\end{align}
with $G_c = i [X, P] - \{ \Psi, \Psi^\dagger \} $. This is a classical statement. Quantization introduces ordering ambiguities, which is evident from the fact that $G_c$ is not traceless. We can fix this by hand by adding a constant part that renders the gauge operator traceless. The result is the constraint
\begin{align}
    \langle \tr G \O  \rangle= 0 \, ,\qquad \text{for all } \O \, .
\end{align}
where 
\begin{align}
    G =  i \left[ X, P \right] - \left\{ \Psi, \Psi^\dagger \right\}  + 2 N I
\end{align}

\subsection{Supercharges}

A SUSY variation of the action will also allow one to derive the supercharges through the equation
\be
\delta S = \int dt \left( \dot{\eta}^\dagger Q - \dot{\eta} Q^\dagger \right) \, .
\ee
The variation obtained by this procedure gives
$$
Q = \left( - i P_{ij} + \frac{\partial W}{\partial X_{ji}}\right) \Psi_{ji}  \quad , \quad Q^\dagger = \left( i P_{ij} + \frac{\partial W}{\partial X_{ji}}\right) \Psi^\dagger_{ji}
$$
These are related to the supercharges in the main text, equation~\eqref{eq:superchargesaa}, by multiplying $Q$ by $-i$ and $Q^\dagger$ by $i$, but this is a matter of definition, or more precisely, a unitary equivalence of the algebra. 

\section{Large$-N$ Harmonic Oscillator}

Here we will give the solution of the quantum harmonic oscillator at large$-N$ for comparison to our results. The solution to these models is reviewed in many places, \textit{e.g.} \cite{Berenstein:2004kk}. We start with the Lagrangian 
\begin{align}
    L = \frac{1}{2} \dot X^2 - \frac{1}{2} X^2
\end{align}
where $X$ and $\dot X$ are matrices in $SU(N)$. The path integral can be transformed into an integral over the eigenvalues at the cost of a Vandermonde determinant 
\begin{align}
    D X \to \Delta(\lambda) \,  DU \,  D\lambda_i \, ,  \qquad \qquad \Delta(\lambda) = \prod_{i \neq j} (\lambda_i - \lambda_j) \, .
\end{align}
The quantization of this theory picks up a determinant in the kinetic term, and we find
\begin{align}
    H = \frac{1}{2} \sum_i \left( - \Delta^{-1} p_i (\Delta p_i) + \lambda_i \right) 
\end{align}
Now, this theory is unitarily equivalent through a transformation.
\begin{align}
    \psi \to \psi' = \sqrt{\Delta} \psi
\end{align}
to another theory with $H' = \sum_i (p_i^2 + \lambda_i^2)/2$. Now the original theory was symmetric in the eigenvalues, but the Vandermonde determinant factor means that the new theory is antisymmetric. The new theory is thus a theory of $N$ free fermions. Its one-particle wave-functions are proportional to hermite polynomials times gaussians, and the full wavefunction can be antisymmetrized using the Slater determinant
\begin{align}
    \tilde \psi = C \det [H^{i}(\lambda_j)] e^{- \sum_i \lambda_i^2 / 2}
\end{align}
where $C$ is a normalization. The determinant depends on the energy level: for $\tilde \psi_{n_1, ..., n_N}$, we have $i \in (n_1, ..., n_N)$, with no $n_l$ being represented more than once due to antisymmetry. Orthogonality of the Hermite polynomials with respect to the Gaussian measure means that many expectation values are very simple to compute: for instance, consider the expectation of a power of $X$,
\begin{align}
    \langle X^n \rangle \ =  \ \int C^2 \det[H(\lambda)]^2 e^{- \sum_i \lambda_i^2 / 2} \sum \lambda_i^n \ = \  \sum_{i \in (n_1, ..., n_N)} \langle \lambda^n \rangle_i
\end{align}
where $\langle ... \rangle_i$ is the expectation value in the eigenstate $H^i(\lambda) e^{-\lambda^2 / 2}$. The powers of $\lambda$ have an explicit form,
\begin{align}
    \langle \lambda^{2k} \rangle_i = \frac{(2k)!}{k! 2^{2k}} \sum_{r}^{\text{min} (k, i)} \frac{(i+k - r)}{(i-r)! (k-r)! r!}
\end{align}

\paragraph{Ground state expectation values} The lowest energy state has the first $N$ numbers occupied, so $n_i = i - 1$. We have 
\begin{align}
    E = \langle H \rangle = \sum_{i = 0}^{N-1}\left( i + \frac{1}{2} \right) = \frac{N^2}{2} \, .
\end{align}
We can compute the first few powers of $X$:
\begin{align}
    \langle X^2 \rangle \ &= \ \frac{1}{2}N^2 \\
    \langle X^4 \rangle \ &= \ \frac{1}{4}N(2N^2 + 1) \\
    \langle X^6 \rangle \ &= \ \frac{5}{8}N^2(N^2 + 2)
\end{align}
To compute expectations involving $P$, we need to write $P$ in the eigenvalue basis. This can be done using $[X_{ab}, P_{cd}] \ = \ i \delta_{ad} \delta_{bc}$ and defining 
\begin{align}
    P_{ab} = U^\dagger_{ax} (\delta_{xy} \pi_x - (1 - \delta_{xy}) J_{xy} / (\lambda_x - \lambda_y) ) U_{yb}
\end{align}
, with $\pi_x = -i \partial_{\lambda_i}$, $J_{xy} = -i U_{xw} \partial_{U_{yw}}$ 
then a slightly tedious calculation shows that 
\begin{align}
     P_{ab} X_{cd} &=  \sum_{x,y,z} U^\dagger_{ax} (\delta_{xy} \pi_x - (1 - \delta_{xy})J_{xy} / (\lambda_x - \lambda_y) ) U_{yb} U^\dagger_{cz} \lambda_z  U_{zd} \\
     &= X_{cd} P_{ab}
     -i \sum_x U^\dagger_{ax}  U_{xb} U^\dagger_{cx}  U_{xd} 
     + i \sum_{x \neq y,z} \frac{\lambda_z}{\lambda_x - \lambda_y} U^\dagger_{ax} U_{xw} U_{yb} \partial_{U_{wy}} U^\dagger_{cz} U_{zd} \\
     %
     %
     %
    &= X_{cd} P_{ab} - i  \delta_{ad} \delta_{bc} \, . 
\end{align}
Using this, we can compute expectation values involving $P$. We find
\begin{align}
    \langle P^n \rangle \ &= \langle X^n \rangle \\
    \langle PX \rangle \ &=  -\frac{i}{2}N^2 \\
    \langle XXPP \rangle \ &= \ \frac{1}{12} N(2N^2 + 1) \\
    \langle XPXP \rangle \ &= \ \frac{1}{12} N(2N^2 -5)
\end{align}
and so on.

\paragraph{expectations with $P$}

\begin{align}
    \langle PX \rangle = \langle \Pi \Lambda \rangle + \langle U^\dagger \tilde J \Lambda U \rangle
\end{align}
Now 
\begin{align}
    \tilde J_{ab} U_{fg} \ = \ - i\frac{ (1 - \delta_{ab}) U_{ac} \partial_{U_{cb}} } {\lambda_a - \lambda_b} U_{fg} \ = \-i  \frac{ (1 - \delta_{ab}) U_{af}  \delta_{bg}  } {\lambda_a - \lambda_b} 
\end{align}
so 
\begin{align}
    \langle U^\dagger_{ab} \tilde J_{bc} \Lambda_{cd} U_{da} \rangle = -i \sum_{a \neq b}\langle  U^\dagger_{ab} U_{be} \partial_{ce} \delta_{cd}  U_{da} \frac{ \lambda_c  } {\lambda_b - \lambda_c}  \rangle =  -i \sum_{b \neq c}\langle  \frac{ \lambda_c  } {\lambda_b - \lambda_c}  \rangle = -\frac{i}{2}N(N-1)
\end{align}
So in total we get 
\begin{align}
    \langle PX \rangle = -\frac{i}{2}N^2
\end{align}

\subsection{Wick contractions}

Expectations involving $P$ are clearly rather cumbersome, but expectations values in the ground state can be quickly computed using Wick contractions:
\begin{align}
    \langle X_{ab}X_{cd} \rangle = \frac{1}{2} \delta_{ad} \delta_{bc} \, , \qquad \langle X_{ab}P_{cd} \rangle = \frac{i}{2} \delta_{ad} \delta_{bc} \, ,\qquad \langle P_{ab}X_{cd} \rangle = \frac{-i}{2} \delta_{ad} \delta_{bc} \\
    \langle P_{ab} P_{cd} \rangle \frac{1}{2} \delta_{ad} \delta_{bc} \, , \quad \ \ \, \qquad \langle \Psi_{ab}\Psi^\dagger_{cd} \rangle =  \delta_{ad} \delta_{bc} \, , \qquad \quad \langle \Psi^\dagger_{ab}\Psi_{cd} \rangle = 0 \, . \qquad \quad 
\end{align}
One can then easily compute ground state correlators by summing over all of these nonzero contractions. We find, for instance
\begin{align}
    \langle \tr \Psi \Psi^\dagger \rangle \ &= \ N^2 \\
    \langle \tr PPXX \rangle \ &= \ -\frac{1}{4} N \\
    \langle \tr \Psi \Psi^\dagger X P \rangle \ &= \ \frac{i}{2} N^3
\end{align}
and so on.

\section*{}
\bibliography{cite.bib}
\bibliographystyle{JHEP.bst}
\end{document}